\documentclass[aps,prfluids,10pt,reprint, onecolumn]{revtex4-2}

\usepackage[normalem]{ulem}
\usepackage{bm}
\usepackage[dvipsnames]{xcolor}
\usepackage{graphicx}
\usepackage{amsmath}
\usepackage[T1]{fontenc}
\usepackage[english]{babel}
\usepackage[utf8]{inputenc}
\usepackage[colorlinks=true, allcolors=blue]{hyperref}
\usepackage{ulem}

\DeclareMathOperator{\St}{St}
\DeclareMathOperator{\Lk}{Lk}
\renewcommand{\vec}[1]{\boldsymbol{\mathbf{#1}}}

\graphicspath{{}}

\usepackage{xcolor}
\usepackage[most]{tcolorbox}
\usepackage{etoolbox}

\definecolor{pvcolor}{RGB}{255,0,0} 
\definecolor{newpccolor}{RGB}{0,0,255}

\usepackage{marginnote}
\reversemarginpar   

\newlength{\pvparindent}
\AtBeginDocument{\setlength{\pvparindent}{\parindent}}

\newtcolorbox{pvblock}[1][]{
  enhanced,
  breakable,
  frame hidden,
  interior hidden,
  boxsep=0pt,
  left=0pt,right=0pt,top=0pt,bottom=0pt,
  borderline west={1.2pt}{-10pt}{pvcolor},
  coltext=pvcolor,
  before skip=\parskip,
  after skip=\parskip,
  parbox=false,
  after upper=\par,
  before=\par\addvspace{\parskip},
  after=\par\addvspace{\parskip},
  before upper={%
    \setlength{\parindent}{\pvparindent}%
    \everypar{\setbox0=\lastbox\indent\everypar{}}%
  },
  overlay unbroken and first={%
    \ifstrempty{#1}{}{%
      \node[
        anchor=north east,
        font=\scriptsize\sffamily,
        text=pvcolor
      ] at ([xshift=-1.4em,yshift=0.2ex]frame.north west) {\shortstack[r]{#1}};
    }%
  },
  overlay first={%
    \ifstrempty{#1}{}{%
      \node[
        anchor=north east,
        font=\scriptsize\sffamily,
        text=pvcolor
      ] at ([xshift=-1.4em,yshift=0.2ex]frame.north west) {\shortstack[r]{#1}};
    }%
  },
}

\newtcolorbox{newpcblock}[1][]{
  enhanced,
  breakable,
  frame hidden,
  interior hidden,
  boxsep=0pt,
  left=0pt,right=0pt,top=0pt,bottom=0pt,
  borderline west={1.2pt}{-10pt}{newpccolor},
  coltext=newpccolor,
  before skip=\parskip,
  after skip=\parskip,
  parbox=false,
  before upper={%
    \setlength{\parindent}{\pvparindent}%
    \everypar{\setbox0=\lastbox\indent\everypar{}}%
  },
  overlay unbroken and first={%
    \ifstrempty{#1}{}{%
      \node[
        anchor=north east,
        font=\scriptsize\sffamily,
        text=newpccolor,
        align=right,
      ] at ([xshift=-1.4em,yshift=0.2ex]frame.north west) {\shortstack[r]{#1}};
    }%
  },
  overlay first={%
    \ifstrempty{#1}{}{%
      \node[
        anchor=north east,
        font=\scriptsize\sffamily,
        text=newpccolor,
        align=right,
      ] at ([xshift=-1.4em,yshift=0.2ex]frame.north west) {\shortstack[r]{#1}};
    }%
  },
}

\begin{document}
\title{Chirality tomography:\\
measuring local helicity from trajectory linking} 

\author{M.~Noseda}
\author{B.~L.~Espa\~nol}
\author{P.~D.~Mininni}
\author{P.~J.~Cobelli}

\affiliation{Universidad de Buenos Aires, Facultad de Ciencias Exactas y Naturales, Departamento de F\'{\i}sica, Ciudad Universitaria, 1428 Buenos Aires, Argentina}
\affiliation{CONICET - Universidad de Buenos Aires, Instituto de F\'{\i}sica Interdisciplinaria y Aplicada (INFINA), Ciudad Universitaria, 1428 Buenos Aires, Argentina}
\affiliation{CNRS-CONICET-UBA, Institut Franco-Argentin de Dynamique des Fluides pour
l’Environnement (IFADyFE), IRL2027, Ciudad Universitaria, 1428 Buenos Aires, Argentina}

\begin{abstract}
We present the first three-dimensional helicity maps of fully developed turbulence obtained through chirality tomography, a Lagrangian voxel-based method that reconstructs helicity density from particle trajectories. Our approach builds on an empirically established relation between helicity and trajectory linking, converting local counts of signed crossings $\mathcal{K}$ into volumetric maps of dimensionless helicity, $\mathcal{H}(\vec{x})$. We demonstrate that the entanglement of particle trajectories, quantified by the mean signed crossing number, provides a robust proxy for helicity, not only at the global scale, but also locally in space and time. Our method can reveal local spatial heterogeneities in helicity and relate them to large-scale flow organization, enabling the reconstruction of spatially resolved chiral structures. Applied to von Kármán experiments and Taylor–Green direct numerical simulations, the method reveals iso-helicity surfaces and coherent chiral features, while time series of $\mathcal{K}$ accurately track the evolution of domain-averaged helicity. The proportionality between $\mathcal{K}$ and $\mathcal{H}$ remains robust across different voxel geometries and different values of particle inertia, but is not held in laminar or time-modulated flows. This study shows that chirality tomography provides a practical helicity diagnostic in turbulent flows, while establishing a direct bridge between trajectory-level topology and a fundamental dynamical invariant of turbulence.
\end{abstract}
\maketitle

\section{Introduction}
\label{sec:S1}

In turbulent flows, quadratic invariants arising from continuous symmetries impose strict constraints on the multiscale transfer of flow quantities, with kinetic energy governing fluctuation intensity across scales and enstrophy controlling small-scale vorticity dynamics in two dimensions. Helicity, defined as the volume integral of the scalar product between the Eulerian velocity  $\vec{u}(\vec{x},t)$ and its curl, $\vec{\omega}(\vec{x},t) = \vec{\nabla} \times \vec{u}$, over the fluid domain $V$,
\begin{equation}
H(t) = \int_V \: \vec{u} \cdot \vec{\omega} \: dV,
\label{eq:H_definition}
\end{equation}
occupies a unique position among these invariant in three dimensions \cite{Moreau1961, Moffatt1969}. Unlike kinetic energy, which is positive-definite, helicity can assume either sign. As a pseudoscalar, $H$ reverses sign under parity transformations, so that a non-vanishing mean reveals broken mirror symmetry, and quantifies the chirality of the flow through the local alignment between velocity and vorticity. The chiral character of helicity hints at a profound topological interpretation, rigorously demonstrated by Moffatt \cite{Moffatt1969}, who established that helicity is proportional to the Gauss linking number of {\it vortex-line pairs} $(\mathcal{C}_i, \mathcal{C}_j)$,
providing a quantitative measure of the average knottedness or linkage of two vortex lines of a flow.  This relation highlights helicity's fundamental role in encoding the topological complexity of fluid flows, irrespective of their laminar or turbulent nature \cite{Moffatt1992, BergerField1984}.

In addition to its topological significance, helicity also constitutes a key dynamical invariant. In three-dimensional inviscid and barotropic flows, helicity conservation \cite{Morrison1998, Salmon1988} follows from the particle-relabeling symmetry of the Euler equations via Noether's second theorem \cite{Padhye1996,Padhye1996b}. In real, viscous flows, however, helicity is not strictly conserved, and its production or destruction is tied to processes such as vortex reconnection and boundary-layer effects \cite{Scheeler2014}.

In high-Reynolds-number turbulence, helicity exerts a strong influence on how energy transfers across scales. Helical triadic interactions become asymmetric, leading to a reduced net flux of energy toward the small scales and can result, when artificially constrained, in an inverse transfer that promotes the growth of large-scale vortical structures \cite{Waleffe1992, Biferale2012}. Direct numerical simulations of isotropic turbulence with imposed helicity confirm modifications to the inertial-range spectral slopes and reduced intermittency compared to non-helical flows \cite{Chen2003}. Such effects are not merely academic: in geophysical and astrophysical contexts, where rotation and stratification often introduce significant helicity, these dynamics underlie cyclone intensification and magnetic dynamo action, respectively \cite{Pouquet_Patterson_1978, Teitelbaum2009}. A deeper understanding of helicity dynamics is therefore essential both for fundamental turbulence theory and for predicting and controlling flows in engineering and natural environments.

Despite its relevance, measuring helicity in laboratory flows remains notoriously challenging, particularly in turbulent regimes. Accurate estimation of $H$ requires knowledge of the vorticity field, which in turn depends on spatial velocity gradients, demanding both high spatial resolution and volumetric coverage. These requirements are difficult to meet in practice. In Eulerian measurements, limited spatial resolution and restricted observation volumes hinder accurate gradient estimation; whereas in Lagrangian approaches that rely on flow-field reconstruction the difficulty is compounded by the indirect nature of the vorticity measurements and the high particle density  required for such detailed reconstructions.

These challenges stem from the lack of dedicated helicity diagnostics and from the limitations of current measurement techniques for velocity-field characterization, which could in principle provide access to helicity. Standard Particle Image Velocimetry (PIV) systems typically yield only two-component velocity fields, and even volumetric or tomographic configurations are restricted to relatively small observation volumes, limiting their ability to resolve the gradients required for vorticity estimation \cite{Raffel2007,Adrian2010}. Three-dimensional Particle Tracking Velocimetry (3D-PTV), in contrast, naturally provides Lagrangian velocity data but offers no direct access to vorticity. Various strategies have been developed to infer this quantity from particle trajectories, ranging from interpolation-based reconstructions \cite{Schanz2016,Schroder2023} to algorithms designed for datasets with sparse trajectory data \cite{Harms2023}. Beyond interpolation-based reconstructions, direct helicity measurements have been demonstrated in carefully controlled settings. Wallace et al. laid the experimental groundwork for helicity density measurements \cite{Wallace1992}; Scheeler and co-workers achieved quantitative, fully three-dimensional measurements in isolated vortex tubes, resolving the link–twist–writhe decomposition and tracking its redistribution during reconnection \cite{Scheeler2014, Scheeler2017}, and Ferraro et al. proposed a local helicity measure designed as a practical diagnostic for turbulent flows \cite{Ferraro2024}. While these studies demonstrate the feasibility of helicity measurement in their respective settings, extending such methodologies to flows with distributed vorticity, such as fully developed turbulence, remains an open challenge.

Such experimental limitations not only render accurate quantification of helicity in complex flows difficult, but also hinder the validation of measurement techniques and theoretical predictions. This highlights the need for benchmark flows where helicity can be tuned and reproduced in a controlled manner. Since mean-flow inhomogeneities serve as a mechanism for helicity production \cite{Yokoi2023}, flows with strong large-scale gradients are prime candidates for such studies. Among these, the Taylor–Green (TG) and von Kármán (VK) flows are of particular interest, as they exhibit qualitatively different chiral properties compared to homogeneous isotropic turbulence (HIT), featuring strong large-scale structures that can act as sources of helicity. The VK turbulent swirling flow consists of two counter-rotating propellers that generate large-scale circulation at high Reynolds number, while TG forcing is a canonical method to produce anisotropic, structured turbulence in numerical simulations. Both configurations have been thoroughly investigated in previous works \cite{Mordant2002,Volk2008,Angriman2020,Angriman2022_2,Angriman2022}, providing well-characterized platforms for studying helicity generation and dynamics.

In an earlier study \cite{Angriman_2021}, a proportionalidy between the mean (dimensionless) helicity $\mathcal{H}$ of a turbulent flow and the average linking number of Lagrangian tracer trajectories, $\mathcal{K}$, was empirically demonstrated. This relation was consistently observed across direct numerical simulations of homogeneous isotropic turbulence (HIT) and Taylor–-Green (TG) flows, as well as in 3D-PTV measurements of laboratory HIT and von Kármán (VK) swirling flows. This showed that the topology of turbulent flows is encoded in the linking of trajectories. Building on this result for helicity as a global quantity, in the present work we introduce chirality tomography, a general approach for reconstructing the spatial distribution of helicity from particle trajectories. Using this technique, we present three-dimensional helicity maps of fully developed anisotropic turbulence, demonstrating its ability to resolve chiral structures in both laboratory and numerical flows. In doing so we establish that the entanglement of particle trajectories, quantified by the average linking number, provides a robust proxy for helicity not only globally but also locally in space and time.

To implement chirality tomography in practice, we developed a Lagrangian, volumetric algorithm that enables the calculation of three-dimensional helicity maps, $\mathcal{H}(\vec{x})$, from normalized, voxel-based counts of apparent trajectory-crossings in two-dimensional projections. Through locally coarse-grained trajectory data, our algorithm reveals spatial heterogeneities and their connection to large-scale flow structures. We validate its performance in turbulent VK and TG flows, using both experiments and direct numerical simulations, and assess robustness across regimes and parameter choices.

Beyond these validations, the ability to reconstruct the three-dimensional helicity density from Lagrangian trajectories motivates four interlinked questions that structure the present work. First, we assess under what assumptions the mean trajectory linking number $\mathcal{K}$ can be related to the helicity $\mathcal{H}$, offering a plausible physical rationale for our tomographic approach; this is the subject of Section~\ref{sec:S2_a}. Second, we test whether our tomography method can reconstruct the spatial field $\mathcal{H}(\vec{x})$ from crossing counts in a statistically stationary turbulent flow, as discussed in Section~\ref{sec:S2_b} and further examined in Section~\ref{sec:S3_a}. Third, we delineate the operational bounds of this method ---how choices of the measuring time interval $\Delta T$, the number $N$ of trajectories considered, the voxel dimensions and shape, and the particle inertia affect the reconstruction accuracy--- as also analyzed in Section~\ref{sec:S3_a}. Finally, we examine the limits of the $\mathcal{K}$–$\mathcal{H}$ relationship itself, identifying flow regimes (laminar or time-modulated) where the proportionality breaks down and probing the underlying causes, which is the matter of Section~\ref{sec:S3_b}. The conclusions of this work are summarized in Section~\ref{sec:S4}.

\section{From topology to tomography}
\label{sec:S2}

In this section we outline the conceptual and methodological framework underlying {\it chirality tomography}. Since our approach relies on the linking of particle  trajectories, which are generically open curves, we begin by clarifying in Section~\ref{sec:S2_z} the meaning of trajectory linking and entanglement in this setting, and by distinguishing its geometric interpretation from the topological notion familiar from closed curves. Section~\ref{sec:S2_a} then presents a rationale linking the average trajectory linking number  $\mathcal{K}$ to the helicity $\mathcal{H}$ in statistically stationary flows, drawing on the topological interpretation of helicity and basic statistical assumptions. This is not intended as a formal derivation, but rather as a physical argument supporting the observed $\mathcal{K}$-$\mathcal{H}$ relation and shedding light on its limitations. In Section~\ref{sec:S2_b} we detail the voxel-based, Lagrangian crossing-counting algorithm that implements this concept into a practical tool for reconstructing spatial helicity maps from particle trajectories. Finally, Section~\ref{sec:S2_c} describes the experimental and numerical datasets used to test and validate the method.

\subsection{Trajectory linking: topology, geometry, and handedness}
\label{sec:S2_z}

The notion of linking has a well-established and intuitive meaning in topology, where it is naturally introduced through the Gauss linking integral \cite{Gauss1833}. For two disjoint (closed or open) oriented space curves $\mathcal{C}_1$ and $\mathcal{C}_2$, with position vectors $\mathbf r_1$ and $\mathbf r_2$ along each curve, the Gauss linking integral is defined as
\begin{equation}
\Lk \left( \mathcal{C}_1, \mathcal{C}_2 \right) = \frac{1}{4 \pi} \int_{\mathcal{C}_1} \int_{\mathcal{C}_2} \frac{\left( \vec{r}_1 - \vec{r}_2 \right)} {| \vec{r}_1 - \vec{r}_2|^3 } \cdot \left( d\vec{r}_1 \times d\vec{r}_2 \right).
\label{eq:Lk_definition}
\end{equation}%
The integrand makes explicit the geometric content of the measure: the kernel $|\vec{r}_1-\vec{r}_2|^{-3}$ emphasizes nearby portions of the curves, while the scalar triple product $(\vec{r}_1-\vec{r}_2)\!\cdot\!(d\vec{r}_1\times d\vec{r}_2)$ encodes their local relative orientation (and hence handedness); the integrals then accumulate these contributions over the full extents of $\mathcal{C}_1$ and $\mathcal{C}_2$. The resulting quantity provides a signed geometric measure of how the two curves wind around one another in three-dimensional space. The Gauss linking integral is well defined for arbitrary (sufficiently regular) space curves, depends smoothly on their relative configuration, and is invariant under global rigid motions. 

For closed curves, this quantity is integer-valued; in this case it is referred to as the Gauss linking number, a topological invariant that classifies links up to continuous deformations without crossings. In this setting, linking admits a clear geometric and physical interpretation: two closed loops are said to be linked if they cannot be separated without cutting one of them; the linking number then counts, algebraically, how many times one loop winds around the other.

An equivalent and particularly intuitive way of accessing the Gauss linking number for closed curves is through the counting of signed crossings in planar projections, a convenient approach employed in related contexts \cite[see][and references therein]{Orlandini1994}. Consider projecting the two curves onto a plane along a generic (non-degenerate) viewing direction, so that their projections intersect at a finite number of transverse crossings. Each crossing is given a sign ($\pm 1$) according to a right-hand convention, based on the local orientations of the projected tangents and on which segment passes {\it over} the other in depth. The algebraic sum of these signs defines the signed crossing number $C(\hat{\vec{n}})$ associated with the viewing direction $\hat{\vec{n}}$. For two oriented closed curves, 
\begin{equation}
\label{eq:closed_crossing_theorem}
C(\hat{\vec{n}}) = 2 \, \mathrm{Lk}(\mathcal{C}_1, \mathcal{C}_2),
\end{equation}%
so that the signed crossing count is independent of $\hat{\vec{n}}$ and coincides (up to a factor) with the integer-valued Gauss linking number in the closed-curve case \cite[see, e.g.,][]{Berger2006}. Thus, for an oriented closed pair, any single generic projection suffices to recover the Gauss linking number.

For open curves, the Gauss linking integral remains perfectly well defined, but it is no longer a topological invariant. Since endpoints may move under continuous deformations without crossings, a knot or link classification analogous to that of closed loops is not available. Instead, the Gauss linking integral provides a continuous geometric measure of mutual winding between the two curves. Its value is no longer constrained to be an integer and varies smoothly with the geometry of the curves, reflecting changes in proximity, orientation, and morphology along their extent. Such a continuous measure is directly relevant for quantifying entanglement in physical systems involving open curves ---e.g. polymeric strands and biomolecules \cite{Foteinopoulou2008, Flapan2000, Laso2009, Sulkowska2012,Shen2024}--- making it a natural tool for finite-time Lagrangian trajectories. Moreover, in the limiting case of nearly closed trajectories (i.e., when the endpoints approach coincidence) the open-curve value approaches the linking number of the corresponding closed link \cite{Panagiotou2013}, so that the closed-loop case is recovered in this limit.

This shift from closed to open curves entails a corresponding change in physical interpretation. Intuitively, for open paths entanglement is not about inseparability but about net mutual helical intertwining: it quantifies whether, as the two oriented trajectories are traversed, one winds around the other with a persistent right- or left-handed bias, rather than through alternating windings of opposite handedness that cancel out. For open curves, entanglement is more naturally associated with handedness than with knottedness, i.e. with the presence or absence of a preferred sense of relative winding. Two trajectories may be trivially separable and yet exhibit a clear right- or left-handed mutual winding when followed along their length; conversely, geometrically intricate configurations may display no net handedness if contributions of opposite orientation balance out. The Gauss linking integral captures precisely this balance, providing a signed measure of entanglement that is sensitive to chirality rather than to topological knotting. In what follows, we use the terms {\it trajectory  linking} or {\it entanglement} interchangeably, in the sense described above, to denote the finite-time net mutual winding of (generically open) particle paths, quantified by $\mathrm{Lk}(\mathcal{C}_1,\mathcal{C}_2)$.

\begin{figure}[t!]
    \centering
    \includegraphics{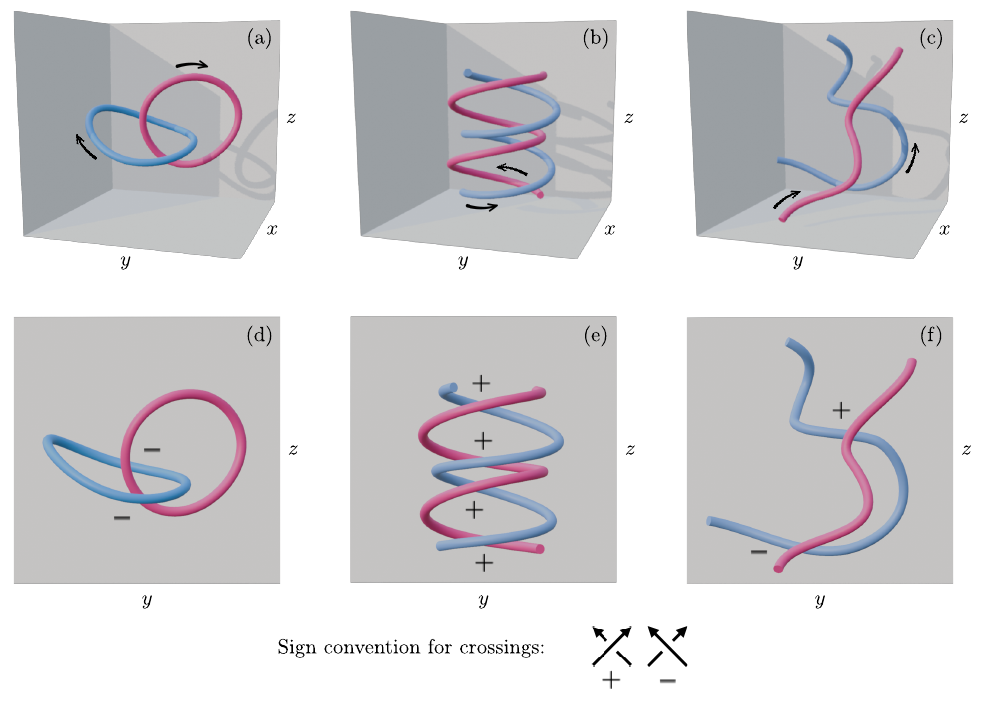}
    \caption{
Illustration of the relation between entanglement and crossings in 2D projections. The figure is organized in two rows and three columns; each column corresponds to a different pair of space curves (e.g., particle trajectories). Panels (a)–(c) (top row) show 3D views, with black arrows indicating the traversal direction along each curve. Panels (d)–(f) (bottom row) display, for the corresponding pairs, an arbitrarily chosen orthographic 2D projection onto the $y$–$z$ plane as viewed from $x>0$ toward the origin. This is one among infinitely many possible projection directions. Whenever the projected curves exhibit an apparent crossing, its sign is assigned according to the convention sketched below the panels. Panel (a) shows the Hopf link (two linked circles). For closed curves, a single generic projection such as that shown in (d) already determines $\Lk$ via the signed crossing number $C(\hat{\mathbf n})$ (see text); in this view both crossings have the same (negative) sign, yielding $\mathrm{Lk}=-1$. Panels (b) and (c) show open curves. In (e), the selected projection presents crossings of equal (positive) sign, consistent with an intertwined configuration. In (f), the apparent crossings compensate with opposite signs, corresponding to a non-entangled pair. For open curves, a single projection is not sufficient: a meaningful estimate of the Gauss linking integral requires averaging crossing information over many projection directions (see Sec.~\ref{sec:S2_z}).}
    \label{fig:diagram}
\end{figure}

This handedness-based interpretation admits a simple formulation in terms of signed crossings in planar projections, for which a rigorous result holds in the open-curve case. For open curves, the signed crossing number obtained from a single projection $C(\hat{\vec{n}})$ generally depends on the viewing direction and does not provide a projection-independent quantity. However, the Gauss linking integral can be recovered by averaging the signed crossing count over all viewing directions \cite{Panagiotou2020}. The average over viewing directions $\langle C(\hat{\vec{n}})\rangle_{\hat{\vec{n}}}$, commonly termed the average signed crossing number, is generally real-valued for open curves. Under standard regularity assumptions on the curves and for generic projection directions, it can be shown that
\begin{equation}
\big\langle C(\hat{\vec{n}}) \big\rangle_{\hat{\vec{n}}} = 2 \, \mathrm{Lk}(\mathcal{C}_1, \mathcal{C}_2).
\label{eq:open_crossing_average_theorem}
\end{equation}
This identity clarifies the role of projection averaging: individual projections may contain many apparent crossings that arise from viewing geometry, but such contributions do not carry a systematic sign bias and therefore cancel upon averaging, while persistent three-dimensional winding manifests as a robust excess of positive or negative (projected) crossings across directions. Note also that, for closed curves, Eq.~\eqref{eq:closed_crossing_theorem} is recovered as a special case of  Eq.~\eqref{eq:open_crossing_average_theorem}, since the signed crossing count becomes independent of the projection direction. A schematic illustration of signed crossings for closed and open curves, and of their behavior under projection averaging, is shown in Fig.~\ref{fig:diagram}. 

In summary, the entanglement of trajectories in three-dimensional space is naturally quantified via the Gauss linking integral between particle paths. This quantity provides a geometric, continuous measure of mutual winding and handedness for open trajectories and is associated with topological linking in the particular case of closed loops. It is generally a real-valued, signed, configuration-dependent quantity that can be estimated directly from Lagrangian data. Moreover, the equivalence between this integral and the average signed crossing number suggests a practical operational estimator based on counting signed crossings in planar projections of pairs of particle trajectories and averaging these counts over viewing directions \cite{Angriman_2021}. In the limit of dense sampling of viewing directions and sufficiently fine curve discretization, this estimator converges to the Gauss linking integral.

For discretely sampled trajectories, the projection-based estimator is preferable to a direct numerical evaluation of the double integral in Eq.~\eqref{eq:Lk_definition} for a number of reasons. Specifically, it avoids numerical sensitivity associated with the near-singular kernel $|\mathbf r_1-\mathbf r_2|^{-3}$ when two trajectory segments come close, is more robust to measurement noise (since direct quadrature involves tangent increments $d\vec{r}$ or local derivatives that can amplify high-frequency fluctuations), and admits efficient implementations with favorable algorithmic scaling. This projection-based formulation provides the basis for the voxelwise crossing-counting methodology introduced later in Sec.~\ref{sec:S2_b}.

With this conceptual framework for interpreting trajectory entanglement established, we now turn to the question of why this geometric measure, when suitably averaged in space and time, becomes quantitatively related to helicity in statistically stationary flows.

\subsection{Helicity and trajectory linking}
\label{sec:S2_a}

In this section we develop the conceptual route that underlies chirality tomography, namely the connection between the local entanglement of Lagrangian trajectories and the helicity of the underlying flow. We start from a dynamical representation of the linking between two tracer trajectories, which separates a bulk vorticity–velocity contribution from a boundary term. We then introduce a local, intensive observable by averaging the two-trajectory linking over all particle pairs sampled within a fixed Eulerian observation volume during a finite time window. Dividing by the window length yields a per-unit-time linking rate. Finally, under statistical stationarity and appropriate local homogeneity assumptions, we relate the ensemble mean of this linking rate to the local helicity density, which provides the basis for reconstructing spatial helicity maps from trajectory data.

Let $\mathcal{C}_1$ and $\mathcal{C}_2$ denote the three-dimensional trajectories of two fluid elements in an incompressible flow with Eulerian velocity $\vec{u}(\vec{x},t)$ and vorticity $\vec{\omega}(\vec{x}, t)$, within a time window given by $t \in \left[0, \Delta T \right)$. The instantaneous position of the tracer particles along these curves is given by $\vec{r}_1(t_1)$ and $\vec{r}_2(t_2)$; their velocities being $\vec{u}_1(t_1) = \vec{u}(\vec{r}_1(t_1), t_1)$ and $\vec{u}_2(t_2) = \vec{u}(\vec{r}_2(t_2), t_2)$. The Gauss linking number for this pair of curves is given by Eq.~\eqref{eq:Lk_definition}, where the vector line elements are related to the flow field through the kinematic relation $d\vec{r}_i = \vec{u}_i \: dt_i$ (with $i = 1, 2$) along the pathlines.

For convenience, let us define
$\phi(\vec{x}_1, \vec{x}_2) = (4 \pi \, | \vec{x}_1 - \vec{x}_2 |)^{-1}$, and $\vec{G}(\vec{x}_1, \vec{x}_2) = (\vec{x}_1 - \vec{x}_2)/(4 \pi \, |\vec{x}_1 - \vec{x}_2|^3)$; both functions related to each other by $\vec{G} = -\vec{\nabla}_{\vec{x}_1} \phi$. In terms of these, the linking number adopts the form
\begin{equation*}
  \Lk(\mathcal{C}_1, \mathcal{C}_2) = \iint \vec{G}(\vec{r}_1(t_1), \vec{r}_2(t_2)) \cdot \big[ \vec{u}(\vec{r}_1, t_1) \times \vec{u}(\vec{r}_2, t_2) \big] \: dt_1 \: dt_2,
\end{equation*}
where the time integrals are taken over the $[0, \Delta T)$ interval. We now introduce the three-dimensional Dirac delta distributions in the space coordinates $\vec{x}_1$, $\vec{x}_2$, defined as $\delta_i(\vec{x}_i, t_i) = \delta^{(3)}(\vec{x}_i - \vec{r}_i(t_i))$. Here $\vec{x}_1$ and $\vec{x}_2$ denote Eulerian spatial coordinates, acting as dummy integration variables when replacing particle positions $\vec{r}_i(t_i)$. By virtue of the identity $\int f(\vec{x}) \: \delta(\vec{x}-\vec{b}) \: d\vec{x} = f(\vec{b})$, valid whenever $\vec{b}$ is in the integration region, the last expression can be recast as
\begin{equation*}
  \Lk(\mathcal{C}_1, \mathcal{C}_2) = \iint \iint \vec{G}(\vec{x}_1, \vec{x}_2) \cdot \big[ \vec{u}(\vec{x}_1, t_1) \times \vec{u}(\vec{x}_2, t_2) \big] \: \delta_1(\vec{x}_1, t_1) \: \delta_2(\vec{x}_2, t_2) \: d^3\vec{x}_1 \: d^3\vec{x}_2 \: dt_1 \: dt_2,
\end{equation*}
with the space integrals taken over a fixed three-dimensional domain $\Omega$ that contains the two trajectories. The triple scalar product in the integrand is equivalent to
$\vec{u}(\vec{x}_1, t_1) \cdot \left\{ \vec{\nabla}_{\vec{x}_1} \times \left[ \phi(\vec{x}_1, \vec{x}_2) \: \vec{u}(\vec{x}_2,t_2) \right]  \right\}$, allowing for integration by parts with respect to $\vec{x}_1$ and leading to
\begin{multline}
  \Lk(\mathcal{C}_1, \mathcal{C}_2)= \frac{1}{4 \pi}\iint \frac{\vec{\omega}(\vec{r}_1(t_1),t_1) \cdot \vec{u}(\vec{r}_2(t_2),t_2)} {|\vec{r}_1(t_1)-\vec{r}_2(t_2)|} \: dt_1 \: dt_2
  \\
  - \frac{1}{4 \pi}\: \iint \oint_{\partial\Omega} \vec{n}\cdot\!\left(\mathbf{u}(\mathbf{x}_1,t_1)\times \frac{\mathbf{u}(\vec{r}_2(t_2),t_2)}{|\vec{x}_1-\vec{r}_2(t_2)|}\right) \delta^{(3)}\left(\vec{x}_1-\vec{r}_1(t_1)\right)\,dS_1\,dt_1\,dt_2,
  \label{eq:Lk_2terminos}
\end{multline}
where the last surface integral is over $\partial \Omega$, the boundary of the chosen integration domain $\Omega$. This expression is a purely kinematic identity for a single pair of tracer particles, with the double time integrals emphasizing that trajectory entanglement is captured asynchronously rather than simultaneously.

The two terms on the right-hand side of Eq.~\eqref{eq:Lk_2terminos} have distinct interpretations. The first one involves a velocity--vorticity inner product, reminiscent of helicity albeit in a different form, weighted by the inverse separation of the trajectories. The second term arises as a surface integral which contributes to the linking number only when a trajectory intersects the integration boundary. In periodic domains this term cancels identically, while in unbounded regions it vanishes provided the integrand decays sufficiently fast. For bounded flows, it is zero under no–slip constraints, and may produce a small but finite contribution for  free–slip conditions. 

In what follows we restrict our attention to cases where the second term in Eq.~\eqref{eq:Lk_2terminos} is negligible; its role on the voxelwise implementation is addressed in Sec.~\ref{sec:S2_b}. The resulting reduced expression for $\Lk$ can then be recast in Lagrangian form by parametrizing trajectories with particle labels. To this end, we introduce the Lagrangian map $\vec{X}(\vec{a},t)$, defined as the position at time $t$ of the fluid element labeled by $\vec{a}$ (for instance, its initial position), so the two trajectories considered so far correspond to $\vec{r}_1(t) = \vec{X}(\vec{a}_1,t)$ and $\vec{r}_2 = \vec{X}(\vec{a}_2,t)$. The corresponding linking number is now
\begin{equation}
\Lk(\vec{a}_1,\vec{a}_2) = \frac{1}{4\pi} \int_{0}^{\Delta T} dt_1
\int_{0}^{\Delta T} dt_2 \:
\frac{\vec{\omega}(\mathbf X(\mathbf a_1,t_1),t_1)\cdot \mathbf u(\mathbf X(\mathbf a_2,t_2),t_2)}
{\left|\mathbf X(\mathbf a_1,t_1)-\mathbf X(\mathbf a_2,t_2)\right|},
\label{eq:Lk_a1a2}
\end{equation}
where the time integrals cover the intervals $[0, \Delta T) \times [0, \Delta T)$.

In order to relate trajectory linking to helicity locally, we now move from a single pair of trajectories to an ensemble of trajectories sampled within a fixed Eulerian observation volume $V$ over a finite time window of length $\Delta T$. For each pair of integration times $(t_1, t_2)$, the average is taken over all particle labels $(\vec{a}_1, \vec{a}_2)$ such that the corresponding trajectory points lie inside $V$, i.e., $\vec{X}(\vec{a}_1, t_1) \in V$ and $\vec{X}(\vec{a}_2, t_2) \in V$. Equivalently, the average retains only the subset of the trajectory ensemble that passes through $V$ during the observation window. In terms of the Lagrangian map, this selection is described by the time-dependent set of labels $A(t) = \vec{X}^{-1}(V, t) = \{ \vec{a} : \vec{X}(\vec{a},t) \in V \}$, so that subsequent averages are performed over $\vec{a}_1 \in A(t_1)$ and $\vec{a}_2 \in A(t_2)$. We emphasize that $V$ is a fixed Eulerian observation volume; since the averaging domain does not move with the flow, the set of labels contributing to the average is necessarily time dependent as particles enter and leave the region.

Building on this, we define the per-unit-time mean linking number in the observation volume $V$ by averaging the two-trajectory linking over all label pairs that contribute at the relevant times. This yields
\begin{equation}
\frac{\Lk_{V,\Delta T}}{\Delta T} =
\frac{1}{V^2\,\Delta T}
\int_{A(t_1)} d^3a_1
\int_{A(t_2)} d^3a_2\;
\Lk(\mathbf a_1,\mathbf a_2).
\label{eq:defLkaverage}
\end{equation}
For each time $t_i$, we perform the change of variables $\vec{x}_i = \vec{X}(\vec{a}_i, t_i)$. Since $A(t_i)$ is the preimage of $V$ under the flow map at time $t_i$, this change of variables maps the integration domain $A(t_i)$ onto the fixed Eulerian volume $V$. For tracer particles in incompressible flows the Lagrangian map is volume-preserving \cite{Holm1999}, with $\det(\vec{\nabla}_{a} \vec{X}) = 1$, so that $d^3 a_i = d^3 x_i$, and the label integrals reduce to spatial integrals over $V$, leading to
\begin{equation}
\frac{\Lk_{V,\Delta T}}{\Delta T} =
\frac{1}{4\pi V^2\,\Delta T}
\int_{V} d^3x_1
\int_{V} d^3x_2
\int_{0}^{\Delta T} dt_1
\int_{0}^{\Delta T} dt_2 \:
\frac{\boldsymbol\omega(\mathbf x_1,t_1)\cdot \mathbf u(\mathbf x_2,t_2)}
{\left|\mathbf x_1-\mathbf x_2\right|}.
\label{eq:aclarada}
\end{equation}
Notice that the volume-conditioned average defined in Eq.~\eqref{eq:defLkaverage} can alternatively be written by extending the label integrals to the full label space and multiplying the integrand by the indicator functions $\mathbf 1_V(\vec{X}(\vec{a}_1,t_1))\,\mathbf 1_V(\vec{X}(\vec{a}_2,t_2))$, which enforce the membership of the two trajectory points in the observation volume at the corresponding integration times. This is strictly equivalent to the restricted integrations in Eq.~\eqref{eq:defLkaverage} and results in the same Eulerian expression given by Eq.~\eqref{eq:aclarada}.

This construction defines a local and intensive measure of trajectory entanglement associated with the observation volume $V$. Normalizing by the window length $\Delta T$ yields a per-unit-time linking rate, which depends on the flow dynamics rather than on the arbitrary duration of the observation interval. Note that this construction implicitly assumes statistical stationarity of the underlying flow, ensuring that the rate converges to a finite limit as $\Delta T \rightarrow \infty$.

In addition, it is customary to introduce a central time $T = (t_1+t_2)/2$ and a time-lag $\tau = t_2 - t_1$. With these definitions, the per-unit-time average linking number over all particle pairs is
\begin{equation}
  \frac{\Lk_{V,\Delta T}}{\Delta T} = \frac{1}{4\pi V^2 \Delta T} \int_V  d^3x_1 \int_V  d^3x_2  \int_{-\Delta T}^{\Delta T}  \int_{|\tau|/2}^{\Delta T - |\tau|/2}
  \frac{\vec{\omega}(\vec{x}_1,T-\tau/2) \cdot \vec{u}(\vec{x}_2,T+\tau/2)}  {|\vec{x}_1-\vec{x}_2|} \: dT \: d\tau,
  \label{eq:5}
\end{equation}
where the subindex $V$ represents the averaging volume, and the subindex $\Delta T$ the average in time. We subsequently focus on the ensemble mean $\langle \Lk_{V,\Delta T}/\Delta T \rangle_\textrm{ens}$ over realizations of the underlying velocity field, under the previously stated statistical stationarity assumption. Introducing the velocity--vorticity correlator $h(\vec{x}_1, \vec{x}_2, \tau) \equiv \langle \vec{\omega}(\vec{x}_1,t) \cdot \vec{u}(\vec{x}_2, t + \tau) \rangle_\textrm{ens}$, the ensemble mean can be written as
\begin{equation}
  \left\langle \frac{\Lk_{V,\Delta T}}{\Delta T} \right\rangle_\textrm{ens} = \frac{1}{4\pi V^2} \int_V  d^3x_1 \int_V  d^3x_2  \int_{-\Delta T}^{\Delta T}
  \left( \frac{\Delta T - |\tau|}{\Delta T} \right)
  \frac{h(\vec{x}_1, \vec{x}_2, \tau)}{|\vec{x}_1-\vec{x}_2|} \: d\tau.
  \label{eq:Lkbar_punto_pre_tauh}
\end{equation}
Here the factor $(\Delta T -|\tau|)$ results from integrating over $T$, since under statistical stationarity the two-time correlator $h(\vec{x}_1, \vec{x}_2, \tau)$ is independent of the central time. We further require that $h$ decays to zero at large time lags and denote its (finite) correlation time by $\tau_h$. Under these conditions, if the observation interval satisfies $\Delta T \gg \tau_h$, the weight $(\Delta T -|\tau|)/\Delta T$ in Eq.~\eqref{eq:Lkbar_punto_pre_tauh} may be set to unity and the $\tau$-integral extended to $(-\infty, \infty)$, up to finite-window corrections of order $\mathcal{O}(\tau_h/\Delta T)$. This results in
\begin{equation}
  \left\langle \frac{\Lk_{V,\Delta T}}{\Delta T} \right\rangle_\textrm{ens} =
  \frac{1}{4\pi V^2}
  \iint_{V\times V}
  \frac{\chi(\vec{x}_1, \vec{x}_2)}{|\vec{x}_1-\vec{x}_2|} \: d^3x_1 \: d^3x_2,
  \label{eq:Lkbar_punto_simplificada}
\end{equation}
where we have defined the time-integrated correlator $\chi(\vec{x}_1, \vec{x}_2)  \equiv \int_{-\infty}^{\infty} \, h(\vec{x}_1, \vec{x}_2, \tau) \: d\tau$.

In order to assess the degree of locality implicit in Eq.~\eqref{eq:Lkbar_punto_simplificada}, it is convenient to rewrite the double spatial integral in terms of the pair separation $\vec{\xi} = \vec{x}_2 - \vec{x}_1$. The key assumption is that the time-integrated correlator $\chi(\vec{x}_1, \vec{x}_2)$ has a finite spatial correlation length $\ell_h$: for $|\xi | \gg \ell_h$, $\chi(\vec{x}_1, \vec{x}_1 + \vec{\xi})$ is small compared to its coincident-point value. Dimensionally, one may anticipate $\ell_h \sim U \tau_h$, where $U$ is a characteristic velocity scale and $\tau_h$ is the (finite) correlation time of the underlying time correlation that defines $\chi$. We stress that $\ell_h$ is introduced here as an effective scale controlling the spatial range of $\chi$, rather than as a sharply defined cutoff.

With this in mind, the $\vec{\xi}$-integration is separated into a near-field region satisfying $|\vec{\xi}|\lesssim \ell_h$, and a far-field region corresponding to $|\vec{\xi}| \gg \ell_h$. This separation is motivated by the structure of the kernel in Eq.~\eqref{eq:Lkbar_punto_simplificada}: the factor $1/|\vec{\xi}|$ gives greater weight to close pairs, so that the dominant contribution is expected to come from   $|\vec{\xi}|$ within the correlation range of $\chi$. The far-field part provides subleading corrections, controlled by the decay of $\chi(\vec{x}_1, \vec{x}_1+\vec{\xi})$ at large separations. In that near-field region we perform a local Taylor expansion of $\chi$ about $\vec{\xi} = \vec{0}$,
\begin{equation*}
\chi(\vec{x}_1, \vec{x}_1 + \vec{\xi}) = \chi(\vec{x}_1, \vec{x}_1) + \vec{\xi} \cdot \vec{\nabla}_{\vec{x}_2} \chi(\vec{x}_1, \vec{x}_2)|_{\vec{x}_2 = \vec{x}_1} + \mathcal{O}(|\vec{\xi}|^2).
\end{equation*}
When inserted into Eq.~\ref{eq:Lkbar_punto_simplificada}, the linear term does not contribute at leading order: the kernel $1/|\vec{\xi}|$ is even under $\vec{\xi} \rightarrow -\vec{\xi}$, and the near-field integration domain can be taken to be isotropic at the scale $\ell_h$, so that the integral of an odd function of $\vec{\xi}$ vanishes. The first non-vanishing correction therefore arises at second order in $\vec{\xi}$, and is small when $\chi$ varies smoothly over distances of order $\ell_h$.

With this leading-order result, $\chi(\vec{x}_1, \vec{x}_1 + \vec{\xi}) \approx \chi(\vec{x}_1, \vec{x}_1)$ becomes independent of $\vec{\xi}$ within the near-field region, and can therefore be taken out of the $\vec{\xi}$-integral. The remaining integral then depends only on the kernel $1/|\vec{\xi}|$ and on the size (and shape) of the neighborhood used to represent the near field, hence it reduces to a purely geometric factor. In three dimensions this factor must scale as $\ell_h^2$, so we can write
\begin{equation*}
  \int_{|\vec{\xi} \lesssim \ell_h} \: \frac{d^3 \xi}{|\vec{\xi}|} = C_\text{geom} \ell_h^2,
\end{equation*}
where $C_\text{geom} = \mathcal{O}(1)$ depends only on the chosen neighborhood shape. For definiteness, choosing a spherical neighborhood of radius $\ell_h$ yields $C_\text{geom} = 2 \pi$. Altogether, we obtain the leading-order local estimate
\begin{equation}
  \left\langle \frac{\Lk_{V,\Delta T}}{\Delta T} \right\rangle_\textrm{ens} \approx
  \frac{\ell_h^2}{2 V} \: \langle \chi(\vec{x}, \vec{x}) \rangle_V
  \label{eq:Lkbar_casi_final},
\end{equation}
where $\langle \cdot \rangle_V$ denotes a spatial average over the domain $V$.

Before turning to the physical interpretation of Eq.~\eqref{eq:Lkbar_casi_final}, we briefly comment on the validity and implications of the approximations involved in the locality reduction. The approximation leading to Eq.~\eqref{eq:Lkbar_casi_final} relies on two distinct locality assumptions. First, it assumes that the time-integrated correlator $\chi(\vec{x}_1,\vec{x}_2)$ has a finite spatial range $\ell_h$, so that contributions from $|\vec{\xi}|\gg \ell_h$ provide only subleading corrections. Second, it assumes that $\chi(\vec{x}_1,\vec{x}_1+\vec{\xi})$ varies smoothly within the near-field region, so that replacing it by its coincident-point value yields the leading-order contribution. In this sense, Eq.~\eqref{eq:Lkbar_casi_final} should be viewed as a leading-order local estimate controlled by the small parameter $(\ell_h/L_\chi)^2$, where $L_\chi$ denotes the typical spatial variation scale of $\chi$.

A further ingredient is the near-field isotropy used to eliminate the linear term in the Taylor expansion. Significant anisotropy at the scale $\ell_h$ (e.g., near solid boundaries or in strongly sheared regions) would in principle generate additional subleading contributions and renormalize the geometric prefactor. Likewise, the specific choice of a spherical neighborhood affects only an $\mathcal{O}(1)$ constant, encapsulated in $C_{\mathrm{geom}}$, but not the $\ell_h^2$ scaling itself. Finally, if $\ell_h$ varies appreciably across the domain, the natural generalization would involve $\langle \ell_h^2(\mathbf{x}),\chi(\mathbf{x},\mathbf{x})\rangle_V$; Eq.~\eqref{eq:Lkbar_casi_final} corresponds to using an effective $\ell_h$ representative of the volume.

Under these assumptions, Eq.~\eqref{eq:Lkbar_casi_final} implies that the mean local linking number is determined by the coincident-point correlation $\chi(\vec{x},\vec{x})$. To interpret this physically, we turn to the underlying velocity–vorticity correlator $h(\vec{x}_1,\vec{x}_2,\tau)$. At coincident points and zero time lag it reduces to the local helicity density; namely: $h(\vec{x},\vec{x},0) = \langle \vec{u}(\vec{x},t)\cdot \vec{\omega}(\vec{x},t)\rangle_\textrm{ens}$. For finite time lags at fixed position, it is natural to introduce the normalized time-correlation function
$C_{\vec{x}}(\tau)=h(\vec{x},\vec{x},\tau)/h(\vec{x},\vec{x},0)$, which quantifies how rapidly velocity and vorticity decorrelate in time. With this representation,
\begin{equation}
  \chi(\vec{x},\vec{x}) = \int_{-\infty}^{\infty} h(\vec{x},\vec{x}, \tau) \, d\tau = \langle \vec{u}(\vec{x},t) \cdot \vec{\omega}(\vec{x}, t)\rangle_\textrm{ens} \int_{-\infty}^{\infty} C_{\vec{x}}(\tau) \, d\tau.
  \label{eq:9}
\end{equation}
The integral on the right hand side defines an effective helicity correlation time $\tau_h^{\text{eff}}(\vec{x})$. Note that this time need not be strictly possitive: if $C_{\vec{x}}(\tau)$ undergoes sign changes, the integral can be strongly reduced and may even vanish, suppressing the linking rate despite a finite instantaneous helicity. In contrast, for the statistically stationary regimes investigated in Sec.~\ref{sec:S3}, the observed proportionality between trajectory linking and helicity is consistent with a finite, non-vanishing $\tau_h^{\text{eff}}$ at the coarse-grained level of our analysis.

If this effective time varies only weakly across the volume, it may be treated as approximately constant, leading to $\langle \chi(\vec{x},\vec{x}) \rangle_V \approx \tau^{\text{eff}}_h \langle \vec{u}(\vec{x},t) \cdot \vec{\omega}(\vec{x},t) \rangle_V$. Substituting into Eq.~\eqref{eq:Lkbar_casi_final} then yields
\begin{equation}
  \left\langle \frac{\Lk_{V,\Delta T}}{\Delta T} \right\rangle_\textrm{ens} \approx \frac{\ell_h^2}{2 V} \, \tau^{\text{eff}}_h \, \langle \vec{u} \cdot \vec{\omega} \rangle_V.
  \label{eq:10}
\end{equation}
This expression makes explicit the sought connection: provided the approximations hold, the ensemble-averaged linking number per unit time is proportional to the volume-averaged helicity density, with proportionality controlled by a geometric factor and an effective helicity correlation time.

To leading order, and assuming statistical stationarity, sufficient homogeneity and isotropy within the volume $V$, and observation windows $\Delta T$ larger than the helicity correlation time; the ensemble mean of the per-unit-time linking number over all particle pairs reduces to a quantity proportional to the mean helicity density. The proportionality involves two factors: a purely geometric prefactor determined by the spatial correlation length $\ell_h$, and an effective helicity correlation time $\tau_h^{\mathrm{eff}}$ that captures the temporal persistence of velocity–vorticity alignment. This result establishes a direct bridge between the geometric measure of trajectory entanglement and the dynamical invariant of helicity. In the next subsection, we describe the geometric algorithm devised to compute linking numbers from particle trajectory crossings within subvolumes (voxels) of the domain, providing the foundation for the tomographic reconstruction technique. This rationale establishes the grounds for using a mean linking number of trajectories $\mathcal{K}$ as a proxy for helicity, setting the stage for the tomographic implementation.

While the above formulation is framed in terms of ensemble averages, many experimental and numerical datasets may instead consist of long temporal records. Accordingly, Appendix~\ref{app:rationale-time-averages} develops a complementary time-average argument under the ergodicity assumption, consistent with the ensemble-based formulation presented in this Section.

In summary, the ensemble-averaged linking rate provides the intensive quantity that remains finite in the long-window limit, and its proportionality to the mean helicity density relies on statistical stationarity and finite decorrelation times. In practice, however, simulations and experiments record the accumulated linking number over a finite interval $\Delta T$. For sufficiently long windows, $\langle \Lk_{V,\Delta T} \rangle_\textrm{ens}$ increases linearly with $\Delta T$, thus providing information physically equivalent to that contained in the intensive rate of Eq.~\eqref{eq:10}.

\subsection{Voxel-based chirality tomography algorithm}
\label{sec:S2_b}

Building on Section~\ref{sec:S2_a}, we introduce a Lagrangian, voxel-based algorithm that leverages the preceding result to compute spatial maps of helicity from local trajectory-crossing statistics. The workflow consists of voxelizing the domain, selecting in-voxel trajectory segments over a prescribed time window, using two-dimensional projections to measure the local three-dimensional entanglement of trajectory pairs, and averaging the resulting measurements across projections. The following paragraphs describe each element in sequence.

The starting point is a set of particle trajectories within a three-dimensional domain $\mathcal{V}$ and sampled at discrete times. The domain is partitioned into a Cartesian grid of $N_x \times N_y \times N_z$ voxels, and a time interval $I_t = [t, t+\Delta T]$ is specified. The choice of $(N_x, N_y, N_z)$ sets the spatial resolution of the tomographic map, and $\Delta T$ defines the temporal coarse-graining used to quantify local winding of trajectory pairs. With this spatial discretization and temporal windowing in place, the steps described below are applied voxelwise within the selected interval.

The algorithm begins by identifying the subset of particle trajectories that co-occupy the voxel at some time during the specified time window, retaining only their in-voxel segments for subsequent analysis. Three-dimensional pairwise winding of trajectories results in oriented crossings in two-dimensional projections, from which we estimate a signed measure of entanglement.


To reduce dependence on the viewing direction, we evaluate this quantity across several projections and average the results. Thus, the degree of linking for trajectory pairs is estimated as the mean sign of crossings, averaged across multiple projections \cite{Kauffman2001}. Note that these crossings are intersections of projected trajectory segments over the time window (path histories), not instantaneous particle encounters. Crossings can also occur between distinct times of the same trajectory; we treat such self-crossings identically, and verified that excluding them yields the same results.

In practice, we draw $P$ distinct projection directions (distributed randomly or quasi-uniformly on the sphere) and project the in-voxel segments onto those planes. In each projection, a crossing is a geometric intersection between two projected segments. Each crossing is given a sign by a right-hand convention with respect to the plane normal: depth order at the intersection fixes which segment is on top, and, together with the directions of motion along the trajectories, determines the orientation. We empirically found that using $P = 2$ projections already provides stable estimates, whose uncertainty diminishes when increasing the number of projections. The results presented in the following section are calculated by averaging over $P = 30$ randomly selected projection directions.

To detect the projected crossings efficiently, we use the classical Bentley–Ottmann sweep-line method \cite{Bentley1979,DeBerg2000}, which reports all intersections with computational cost $\mathcal{O}((s+c)\log s)$, where $s$ is the number of projected segments and $c$ the number of crossings. The method also returns the identities of the intersecting segments, allowing each two-dimensional event to be mapped back to its three-dimensional trajectory pair and the crossing sign to be determined unambiguously.

For the $p$-th projection of trajectories in voxel $V$ during the interval $I_t$, we compute the normalized crossings $K_p$ as the mean of the crossing signs for that plane: $K_p = M_p^{-1}\sum_{i=1}^{M_p} \sigma_i$. Here $M_p$ is the total number of apparent crossings in this projection and $\sigma_i = \pm 1$ represents the sign of the $i$-th crossing. We then define the mean linking number, denoted as $\mathcal{K}_{V,\Delta T}$, as the mean of $K_p$ over all $P$ projections,
\begin{equation}
    \mathcal{K}_{V, \Delta T} = \frac{1}{P} \sum_{p=1}^P \, K_p.
    \label{eq:average_lk}
\end{equation}
Algorithmically, this quantity serves as our entanglement proxy built from projected crossings: over a set of random directions we project the trajectories, tally in each projection the pairwise algebraic (signed) and total (unsigned) crossing counts, and define $\mathcal{K}_{V, \Delta T}$ as the projection-weighted fraction, with weights proportional to the total counts.

Apart from voxelwise localization and adopting the Bentley–Ottmann method for efficient intersection detection, the procedure outlined here to estimate the mean linking number $\mathcal{K}_{V,\Delta T}$ from trajectories closely follows the method used in \cite{Angriman_2021}. The code developed for the analysis of trajectories described in this section is made available at \cite{NosedaZenodo2025} (see Section~\ref{sec:data}).

To streamline notation, in the following we suppress explicit subscripts and write simply $\mathcal{K}$ when the context is unambiguous, with $(V,\Delta T)$ left implicit and reintroduced only when clarity demands.

It is worth emphasizing that the voxelwise estimator described above implements the reduced (bulk) form of Eq.~\eqref{eq:Lk_2terminos}, i.e. it does not explicitly account for the surface contribution. As mentioned in Sec.~\ref{sec:S2_a}, that surface term acts as a boundary-crossing correction, activated only by trajectory segments intersecting voxel faces. In statistically stationary regimes and for observation windows $\Delta T$ longer than the relevant correlation times, entry and exit events across voxel faces are expected to balance in the long-time average, so that the net effect of this correction becomes subleading compared with the volumetric term. Moreover, when the voxel linear size exceeds the relevant spatial correlation length $\ell_h$, the relative magnitude of boundary contributions is suppressed by a geometric factor scaling as $\ell_h/V^{1/3}$ (up to an $\mathcal{O}(1)$ shape factor depending on voxel geometry). The robustness of the reconstructed maps and correlations under voxel size/geometry variations (presented in Sec.~\ref{sec:S3}) provides an assessment that residual boundary-crossing effects do not measurably bias the leading-order relation for the parameter ranges considered here.

\subsection{Experimental and numerical datasets}
\label{sec:S2_c}

\begin{figure}[t]
    \centering
    \includegraphics{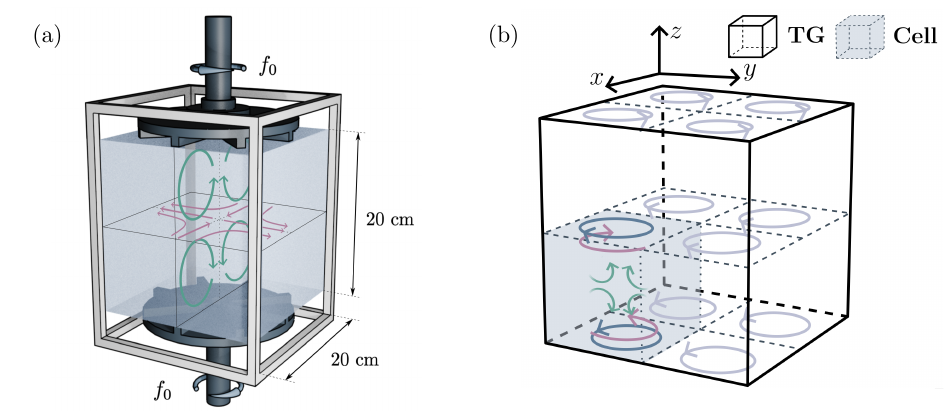}
    \caption{Schematics of the experimental and numerical configurations.
        (a) Von K\'arm\'an experimental setup; $f_0$ denotes the rotation frequency of the impellers. The arrows inside the cell illustrate the principal features of the mean large-scale flow.
        (b) Geometry used for the Taylor-Green simulations.
        The periodic domain, a cube of side $(2\pi L_0)$, contains eight subcells of volume $(\pi L_0)^3$, each of which hosts large-scale structures reminiscent of those in the VK flow. This correspondence is illustrated by the highlighted subcell. Figure adapted from \cite{Espanol2025}.}
    \label{fig:datasets_schemes}
\end{figure}

In this study we consider two complementary Lagrangian datasets of particle trajectories spanning laboratory and numerical turbulence. Detailed parameters (forcing, resolution, particle properties, Reynolds numbers, as well as acquisition or simulation settings) are given below. These configurations are chosen based on the similarities observed between the two flows in previous studies \cite{Angriman2020, Angriman2022_2}.

The VK experimental setup consists of an acrylic tank of square cross-section with two counter-rotating impellers surrounding a $20 \times 20 \times 50$ cm$^{3}$ volume. Each impeller drives a disk with eight straight blades that generate two large-scale counter-rotating anisotropic structures with strong shear in the midplane. These produce horizontal structures with larger correlation length than their vertical counterparts. Fig.~\ref{fig:datasets_schemes}(a) presents a schematic of the main parts of the VK setup. Distilled water from a double-pass reverse-osmosis system is used as the working fluid, which is kept at 25°C using two coils connected to a chiller, placed behind the blades (not shown in the diagram). The impellers are run at $f_0 = 50 \, \rm{RPM}$ (equivalently, $0.83$~Hz), yielding a Kolmogorov timescale of $\tau_\eta = (\nu / \varepsilon)^{1/2} = 0.011 \, \rm{s}$, where $\nu$ is the kinematic viscosity of water and $\varepsilon$ is the average energy dissipation rate.

The numerical datasets were obtained from a direct numerical simulation (DNS) of the incompressible Navier--Stokes equations
\begin{equation}
    \partial_t \vec{u} + \vec{u} \cdot \vec{\nabla} \vec{u} = - \vec{\nabla} p + \nu \nabla^2 \vec{u} + \vec{F}, \ \ \ \nabla \cdot \mathbf{u} = 0,
\end{equation}
where $\vec{u}$ is the velocity field, $p$ is the pressure per unit mass density, and $\vec{F}$ represents the external mechanical forcing. The simulation was performed in a $(2\pi L_0)^3$-periodic cubic domain, using the GHOST code, a massively parallel pseudospectral solver for partial differential equations in periodic domains \cite{Mininni2011, Rosenberg2020}. In this setting, the characteristic length was set to $L_0 = 1$. The Taylor-Green (TG) simulation uses a structured forcing that generates large-scale counter-rotating vortices in each $(\pi L_0)^3$ subvolume \cite{Taylor1937},
\begin{equation}
    \vec{F} = F_0 \left[ \sin(k_F x) \cos(k_F  y) \cos(k_F  z) \: \hat{x}  - \cos(k_F  x) \sin(k_F  y) \cos(k_F  z) \: \hat{y} \right],
\end{equation}
when the forcing wavenumber is set to $k_F = 1$. We show a diagram of this flow in Fig.~\ref{fig:datasets_schemes}(b). The flow is discretized in $768^3$ grid points and evolved for over $15 \, T_0$. The simulation has a turbulent steady state with well-resolved dissipation scales. 

A key dimensionless parameter used to quantify the intensity of turbulent fluctuations is the Taylor-scale Reynolds number,
\begin{equation}
    \text{Re}_\lambda = \sqrt{15\, \frac{U^4}{\nu \varepsilon}}.
\end{equation}
The Taylor-scale Reynolds numbers for our experimental and numerical datasets are $\text{Re}^{\rm{VK}}_\lambda = 241$ and $\text{Re}^{\rm{TG}}_\lambda = 351$, respectively.

Particles in both our experimental and numerical datasets are characterized by their Stokes number, $\St = \tau_p/\tau_\eta$, which measures the particles' response time $\tau_p$ in terms of a characteristic flow time scale, here taken as the Kolmogorov timescale $\tau_\eta$. The VK is seeded with neutrally buoyant polyethylene particles of $250$–$300 \, \mu$m in diameter, which yields a value of $\St = 0.78$. The particles are illuminated from behind using two LED panels and two high-speed cameras are used to track particles at 1 kHz within a central cubic region of $(16 \, \rm{cm})^{3}$ volume using 3D-PTV. For the present study, 20 experimental realizations were recorded, each spanning $1.3 \, T_0$, where $T_0$ is the integral timescale computed as $T_0 = L_0/U$, using $U$ as the root mean square (RMS) velocity of the particles using only the $x$ and $y$ components (due to anisotropy), and $L_0$ as a characteristic length of the flow, taken as the disk diameter $L_0 = 0.19 \, \rm{m}$. This procedure yielded $\mathcal{O}(10^4)$ trajectories per experimental run, with a mean of 205 particles per frame and an average trajectory lifespan of $0.3 \, T_0$. Further experimental details can be found in \cite{Angriman2020, Espanol2025}. 

In the simulations, particle trajectories are generated for two Stokes numbers. The first dataset uses particles with $\St = 0.76$, selected to match the experimentally relevant regime described above and thus provide the most directly comparable numerical counterpart of the measured observable. A second dataset, with $\St = 8.89$, is considered for a different purpose: it probes a markedly more inertial regime and tests the persistence of the trajectory linking--helicity correlation under stronger inertial effects, which may modify the sampling of the flow along particle trajectories. For reference, the tracer-limit results ($\St = 0$) are reported in \cite{Angriman_2021}.

In each numerical particle dataset, $10^6$ neutrally buoyant one-way-coupled particles were evolved using the dynamical models detailed below. Firstly, for particles with $\St = 0.76$, we use the Maxey-Riley approximation \cite{Maxey1983,Gatignol1983,Auton1988}, neglecting both the Fax\'en corrections \cite{Happel1983} and the Basset-Boussinesq history term \cite{Boussinesq1885,Basset1888},
\begin{equation}
    \dot{\mathbf{x}}_p = \mathbf{v}_p, \ \ \  \dot{\mathbf{v}}_p = \frac{1}{\tau_p}\big[\mathbf{u}(\mathbf{x}_p,t)-\mathbf{v}_p\big].
\end{equation}
where $\vec{x}_p$ and $\vec{v}$ stand for the particle's position and velocity, respectively. For inertial particles characterized by $\St = 8.89$, we employed a nonlinear-drag model, as done in \cite{Wang1993, Angriman2022}, namely
\begin{equation}
    \dot{\vec{x}}_p = \mathbf{v}(t), \ \ \ \dot{\vec{v}} = \frac{1 + 0.15 \, \rm{Re}_p^{0.687}}{\tau_p}\left[ \vec{u}(\vec{x}_p, t) - \mathbf{v}(t) \right],
\end{equation}
where $\rm{Re}_p = \sqrt{18 \tau_p/\nu}|\vec(\vec{x}_p, t) - \vec{v}(t)|$.

\section{Results}
\label{sec:S3}

\subsection{Helicity maps of turbulent flows}
\label{sec:S3_a}

\begin{figure}[t!]
    \centering
    \includegraphics[width=\linewidth]{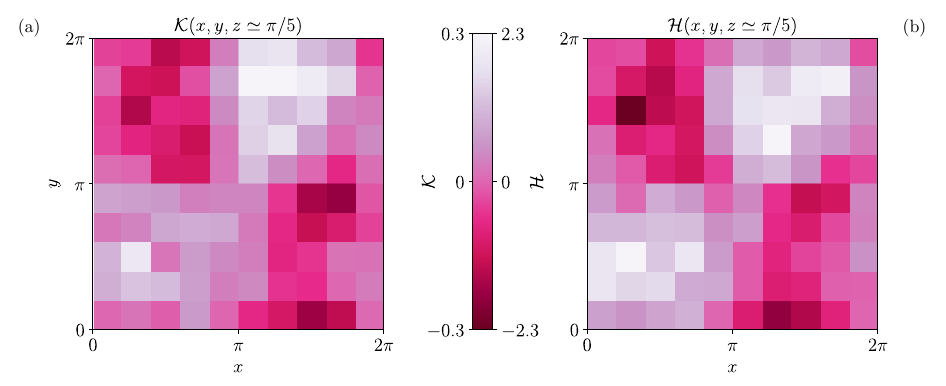}
    \caption{Tomographic slice of the Taylor--Green turbulent flow: reconstructed $\mathcal{K}$ and ground-truth $\mathcal{H}$. (a) Time-averaged mean linking number, computed from $N = 10^4$ trajectories and mapped onto $10^3$ cubic voxels; slice on a plane parallel to $x$--$y$ at $z \simeq \pi/5$. (b) Dimensionless helicity obtained from the DNS fields, coarse-grained to the same voxel geometry and averaged over the same interval; slice at the same location. The characteristic four-cell organization of the flow is visible in both panels.}
    \label{fig:cross}
\end{figure}

Our first objective is to assess whether chirality tomography, based on local trajectory entanglement, can capture the underlying chiral structures of turbulence. To address this question, we begin with the Taylor--Green dataset, where Eulerian velocity fields are available and vorticity can be computed explicitly, providing ground-truth maps of the non-dimensional helicity, 
\begin{equation}
    \mathcal{H} = \frac{H}{U^{2} L^2},
\end{equation}
where $L$ and $U$ are taken as the characteristic length of the domain and the RMS flow velocity, respectively. This makes the TG dataset an ideal benchmark to validate our method, by comparing helicity distributions with the corresponding maps of the mean linking number $\mathcal{K}$.

To construct local tomographic fields, we partition the observation volume into a regular Cartesian grid of $N_x \times N_y \times N_z$ voxels. The triplet $\mathbf{N} = (N_x, N_y, N_z)$ uniquely determines the voxel aspect ratio. We first consider the TG domain and subdivide it into $\mathbf{N}=(10,10,10)$ cubic voxels, and averaged trajectory crossings over consecutive non-overlapping time windows of duration $\Delta T = T_0$. To mimic experimental conditions, only a subsample of $N = 10^4$ particles was retained. This procedure yields spatial maps of the average signed crossing number $\mathcal{K}$. The ground-truth helicity field $\mathcal{H}(\vec{x})$ is obtained at full simulation resolution, but for a fair comparison it is coarse-grained to the same voxel size used for the trajectory counts, and averaged in time over the same time windows of length $\Delta T$. For clarity of presentation, Fig.~\ref{fig:cross} displays a two-dimensional slice of these fields on a plane parallel to the $x-y$ axis at height $z \simeq \pi/5$: panel (a) presents the reconstructed distribution of $\mathcal{K}$, while panel (b) shows the corresponding coarse-grained helicity. Despite the relatively low resolution, the four characteristic TG cells are clearly discernible, suggesting consistency between the reconstructed field and the flow organization. Beyond this illustrative slice, the quantitative comparison is performed over the entire set of $10^3$ cubic voxels spanning the domain. In this full-volume analysis, the Pearson correlation coefficient between $\mathcal{K}(\vec{x})$ and $\mathcal{H}(\vec{x})$ reaches $R = 0.82$, indicating that the voxel-wise correspondence is strong. This result provides evidence that trajectory-based tomography captures the large-scale chiral structures of the flow even when using only a limited number of particles and coarse voxel resolution.

\begin{figure}[t!]
    \begin{minipage}[t]{0.48\textwidth}
      \raggedright
      (a)\\
      \includegraphics[width=\linewidth]{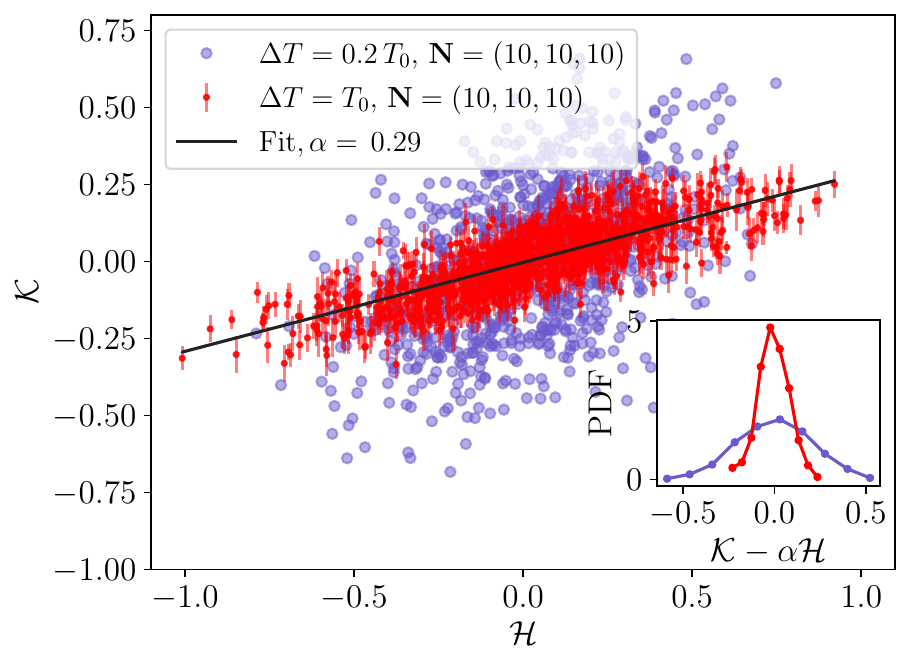}\hspace{0.5cm}
    \end{minipage}\hfill
    \begin{minipage}[t]{0.48\textwidth}
      \raggedright
      (b)\\
      \includegraphics[width=\linewidth]{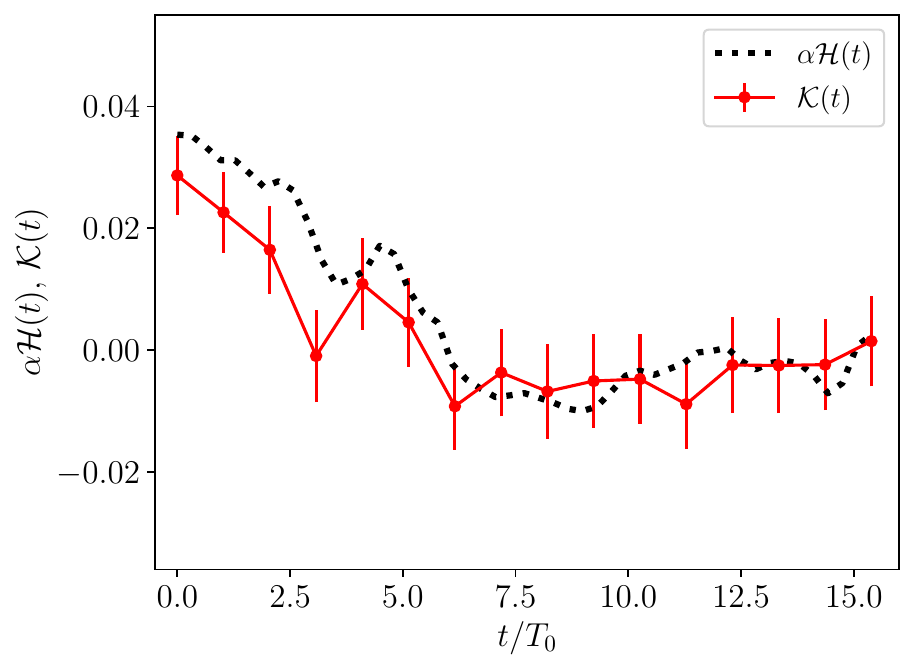}
    \end{minipage}
    \begin{minipage}[t]{0.48\textwidth}
      \raggedright
      (c)\\ \hspace*{0.635cm}%
 \includegraphics[width=0.925\linewidth]{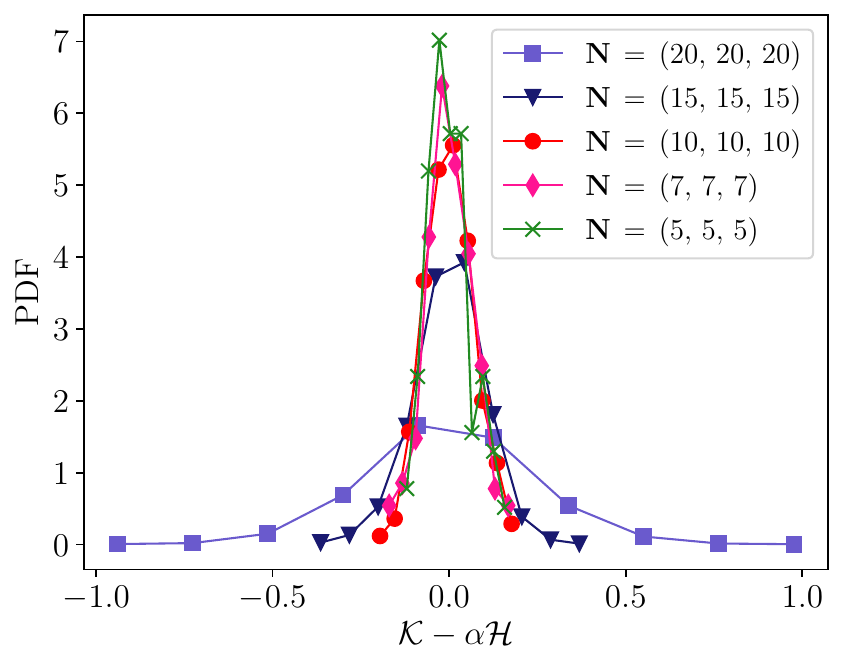}
    \end{minipage}\hfill
    \begin{minipage}[t]{0.48\textwidth}
      \raggedright
      (d)\\
      \includegraphics[width=\linewidth]{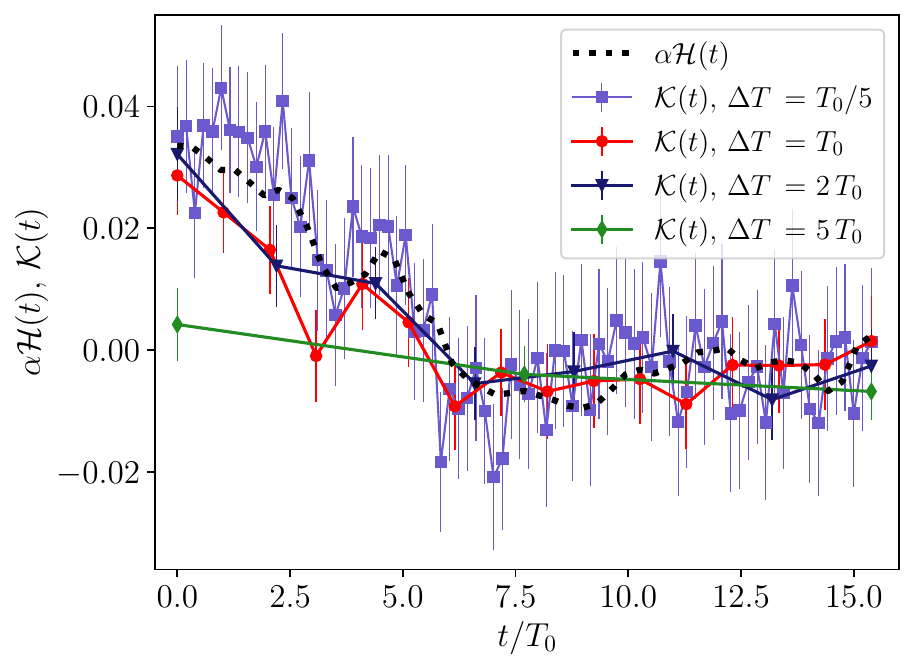}
    \end{minipage}
    \caption{Mean linking number as a proxy for helicity in the TG flow. (a) Time-averaged values of $\mathcal{K}$ obtained from a subdivision of the domain into $10^3$ cubic voxels. Red symbols correspond to $N=10^4$ trajectories observed for a time window of length $\Delta T = T_0$, using $20$ random projections; error bars denote the standard error across ten independent subsamples. Blue symbols show the same analysis with shorter histories, $\Delta T=0.2 \, T_0$, and $N=10^3$ particles. Insets display the probability density functions of the residuals. (b) Spatial average of $\mathcal{K}$ computed with the same parameters as in the red dataset of the left panel, compared with the domain-averaged dimensionless helicity $\mathcal{H}$ (dotted black line), rescaled by the fitted proportionality constant. (c) Spread of the PDF of residuals for different spatial resolutions, $\mathbf{N}$. The deviation increases as the number of voxels grows, since the available statistics are finite. Conversely, no significant reduction of the spread is observed when using $\mathbf{N} = (10, \, 10, \, 10)$ voxels or fewer. (d) Comparison of $\mathcal{K}(t)$ obtained with different temporal windows. Shorter time windows provide improved temporal resolution, as the entanglement of trajectories is averaged over shorter intervals; however, the associated uncertainty increases due to reduced statistics, even though the mean value remains close to the non-dimensional helicity.}
    \label{fig:k_vs_h}
\end{figure}

To assess the statistical relation between the mean trajectory-linking number and helicity, we first consider voxel-wise values in the TG dataset. Fig.~\ref{fig:k_vs_h}(a) shows $\mathcal{K}$ against the dimensionless helicity $\mathcal{H}$, using the same dataset employed in Fig.~\ref{fig:cross}, i.e., a subsample of $N = 10^4$ particles. A weighted linear fit yields a slope of
$\alpha_{\rm{TG}} = (0.289 \pm 0.007)$ with a near-zero intercept, comparable to the value reported previously in \cite{Angriman_2021}. For comparison, we also repeated the analysis with fewer particles ($N = 10^3$) and shorter averaging windows ($\Delta T = 0.2 \, T_0$). As expected, this dataset exhibits larger scatter, as evidenced in the inset by the broader distribution of residuals from the fit, but still displays correlation.

A natural complementary question is whether the relation between $\mathcal{K}$ and $\mathcal{H}$ also holds in time. To examine this, we focus on the same dataset represented in red in Fig.~\ref{fig:k_vs_h} (a). Averaging over all voxels yields a time-dependent domain-wide measure $\mathcal{K}(t) \equiv \langle \mathcal{K}_{V,\Delta T} \rangle_V(t)$, resolved at the temporal scale of the averaging window $\Delta T$. The resulting time series is displayed in Fig.~\ref{fig:k_vs_h}(b), where the proxy $\mathcal{K}(t)$ is compared with the ground-truth evolution of $\mathcal{H}(t)$, rescaled by $\alpha$, obtained from the DNS fields under equivalent conditions (time-windowing, spatial coarse-graining). The two time series follow each other closely within the error bars, indicating that $\mathcal{K}$ not only recovers the spatial organization of helicity but also its temporal evolution at the resolution set by the measurement interval. To verify, we run these tests on a dataset of tracer particles and we found no significant differences with the shown results.

The choice of spatial and temporal resolution is primarily constrained by the amount of locally available trajectory-crossing statistics. Increasing either the spatial or temporal resolution requires resolving a larger number of crossings within each voxel or time window. Conversely, reducing the resolution amounts to averaging the helical dynamics over larger spatial or temporal scales, effectively acting as a low-pass filter that smooths out small-scale chiral structures. In this regime, longer integration times are required to accumulate statistically meaningful crossing counts. Even in the presence of sufficient statistics, the resolution cannot be increased arbitrarily: a physical upper limit is set by the Kolmogorov scale, as the linear relation between crossings and helicity is established for statistically stationary turbulent flows. Fig.~\ref{fig:k_vs_h}(c) illustrates the effect of varying the spatial resolution. Increasing the resolution to $\mathbf{N}=(20,\,20,\,20)$ leads to a marked increase in the spread of the residuals due to insufficient statistics within individual voxels. Conversely, lowering the resolution below $\mathbf{N}=(10,\,10,\,10)$ does not significantly reduce the spread, as this coarse graining blends regions with different local dynamics due to the inhomogeneity of the flow, leading to average different entanglement dynamics and yielding an increased variability. Fig.~\ref{fig:k_vs_h}(d) shows the effect of varying the temporal window $\Delta T$. Shorter windows yield a time evolution of $\mathcal{K}$ that more closely follows the non-dimensional helicity value; however, the reduced number of crossings within each window leads to a larger relative uncertainty. We also notice that when the time window is too large, the small scale fluctuations in the helicity of the flow are expectedly loss in the variations of $\mathcal{K}$.

\begin{figure}[t!]
    \centering
    \includegraphics[width=0.9\linewidth]{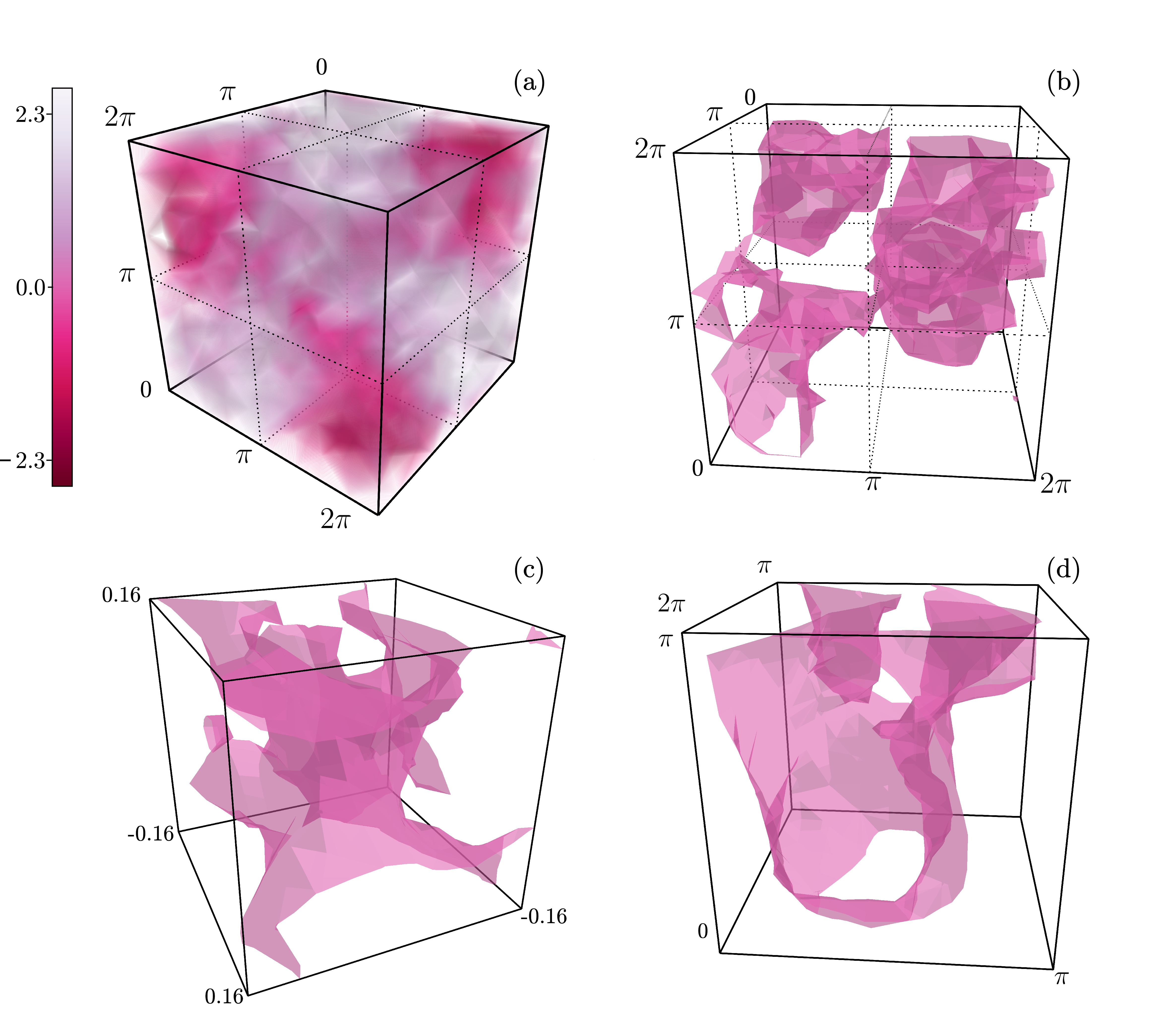}
    \caption{Three-dimensional tomographic reconstructions of helicity in Taylor--Green (numerical) and von K\'arm\'an (experimental) turbulent flows. (a) Volumetric rendering of the reconstructed field $\mathcal{K}/\alpha$ for the TG dataset. Individual subcells can be clearly distinguished, with helicity alternating sign between neighboring regions, a characteristic feature of the TG flow. (b) Isosurfaces at $\mathcal{H}=0.4$ in the full TG domain. (c) Isosurfaces at $\mathcal{K}=0.07$ for the VK experiment in the full observation volume (units in meters). (d) Zoom of the isosurfaces shown in (b), restricted to the subcell $[\pi, \, 2\pi] \times [0, \, \pi] \times [0, \, \pi]$. }
    \label{fig:tomographic-results}
\end{figure}

Fig.~\ref{fig:tomographic-results} presents the full three-dimensional tomographic reconstruction of the helicity fields through chirality tomography. Panel (a) shows the volumetric rendering of $\mathcal{K}$ for the TG dataset, from which the slice in Fig.~\ref{fig:cross} was extracted. Panel (b) displays the $\mathcal{H}$–isosurfaces for $\mathcal{H} = 0.4$ in the full domain, while (c) zooms into a single TG cell to highlight the local chiral structures. Panel (d) presents $\mathcal{K}$–isosurfaces for the VK dataset, enabling a direct visual comparison across flows. The large-scale handed regions, vortex-like lobes, and their spatial organization are consistent with those found in tests we performed on low-order helical models of both flows \cite{Berning2023,Espanol2025}. In short, panels (b-d) showcase the isosurfaces of helicity, which capture the chiral structures only through linking trajectory information.

\begin{figure}[t!]
    \begin{minipage}[t]{0.48\textwidth}
      \raggedright
      (a)\\
      \includegraphics[width=\linewidth]{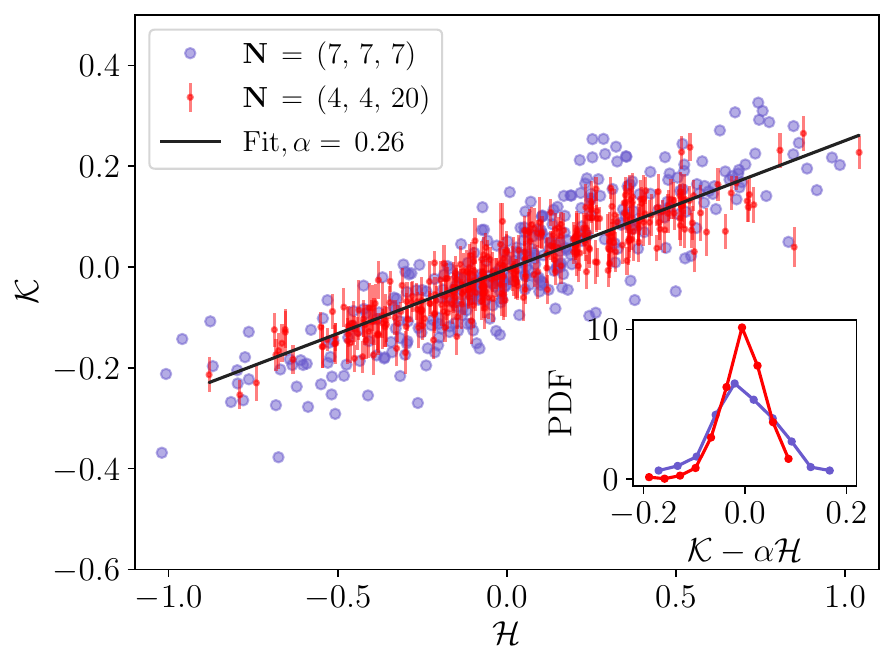}\hspace{0.5cm}
    \end{minipage}\hfill
    \begin{minipage}[t]{0.48\textwidth}
      \raggedright
      (b)\\
      \includegraphics[width=\linewidth]{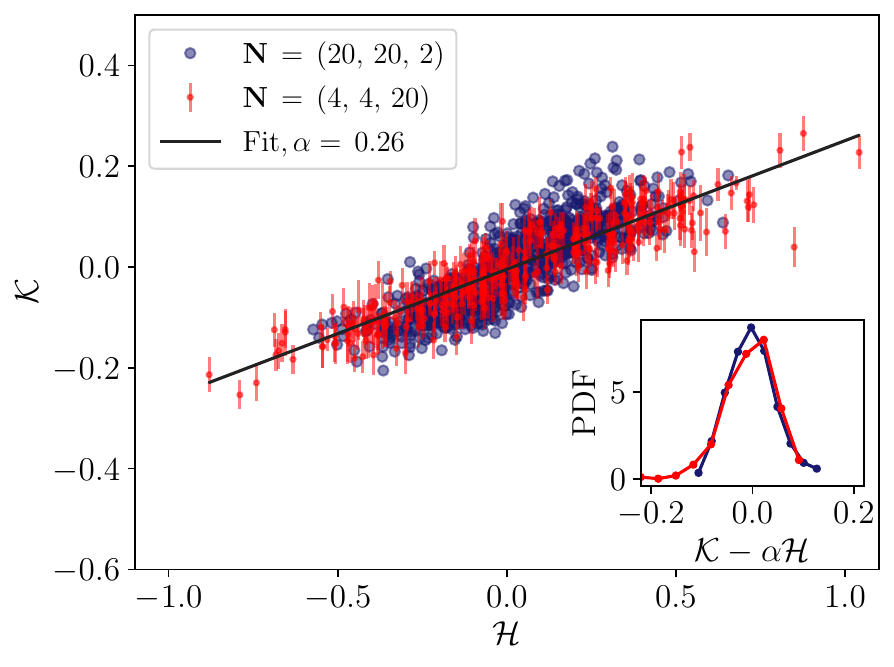}
    \end{minipage}
\caption{Effect of voxel geometry on the $\mathcal{K}$--$\mathcal{H}$ relation in the TG dataset. (a) Time-averaged values of the mean linking number for the isotropic subdivision $\mathbf{N}=(7,7,7)$ (large blue circles) and the laterally stretched subdivision $\mathbf{N}=(4,4,20)$ (small red circles). Both cases use $\Delta T = T_0$ for $N = 10^4$ trajectories. Insets show the probability density functions of the residuals. (b) Comparison between the laterally stretched subdivision $\mathbf{N}=(4,4,20)$ (small red circles) and the column-like subdivision $\mathbf{N}=(20,20,2)$ (dark blue circles).}
    \label{fig:k_vs_h_geometry}
\end{figure}

We next examine how voxel number and geometry affect the relation between trajectory-linking number and dimensionless helicity. For the present analysis we consider three representative subdivisions of the complete volume: an isotropic case, $\mathbf{N}=(7,7,7)$ (cubic voxels); a laterally stretched case, $\mathbf{N}=(4,4,20)$ (voxels five times wider than tall); and a column-like case, $\mathbf{N}=(20,20,2)$ (voxels ten times taller than wide). Fig.~\ref{fig:k_vs_h_geometry}(a) compares the $\mathcal{K}$--$\mathcal{H}$ relation for the isotropic and laterally stretched subdivisions (large blue and small red circles, respectively); the solid black line shows the linear fit to the latter dataset. A slight reduction in scatter is observed, which we attribute to the laterally stretched voxels being better matched to the flow's anisotropic organization and thus capturing its dominant structures more consistently. In contrast, Fig.~\ref{fig:k_vs_h_geometry}(b) compares the laterally stretched and column-like subdivisions (small red and dark blue circles, respectively); the solid black line again shows the linear fit to the laterally stretched dataset. The increased vertical averaging in the column-like case smooths out local fluctuations and reduces the occurrence of extreme values in $\mathcal{H}$. These results illustrate that voxel geometry modulates the balance between resolution and sensitivity in the $\mathcal{K}$--$\mathcal{H}$ trend, while preserving the underlying correlation. 

\begin{figure}[t!]
    \centering
    \includegraphics[width=0.4815\linewidth]{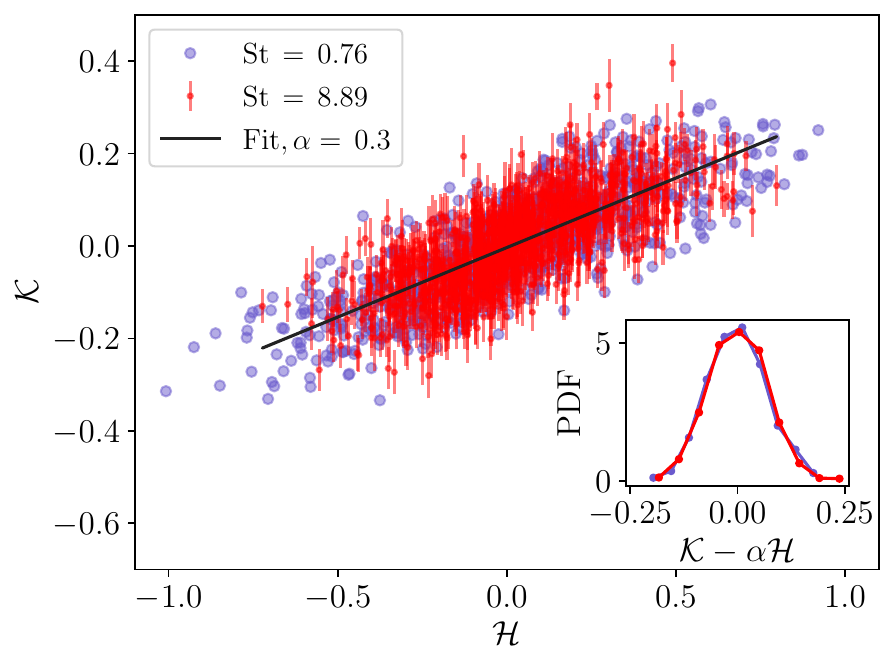}
    \caption{Robustness of the $\mathcal{K}$--$\mathcal{H}$ relation with particle inertia in the TG dataset. Time-averaged values of $\mathcal{K}$ over a subdivision of $10^3$ cubic voxels, computed for $\Delta T = T_0$ and considering $N = 10^4$ trajectories, for particles with $\St = 0.76$ (large blue circles) and $\St = 8.89$ (small red circles). Insets: probability density functions of the residuals from the corresponding linear fits.}
    \label{fig:K_vs_h_stokes}
\end{figure}

An aspect not explored so far is the role of particle inertia in the $\mathcal{K}$--$\mathcal{H}$ relation. Just as energy exhibits a direct cascade, helicity also undergoes a forward transfer across scales. In homogeneous and isotropic turbulence, its spatial (wavenumber) spectrum scales as $k^{-5/3}$ \cite{Chen2003}, which implies that helicity is predominantly concentrated at large scales. One thus expects trajectory entanglement to be primarily governed by these large-scale motions, so that particle inertia ---acting as a low-pass filter--- should not significantly affect the result. A direct test is provided by the inertial-particle dataset at $\St = 8.89$ (Sec.~\ref{sec:S2_c}). The $\mathcal{K}$--$\mathcal{H}$ analysis is carried out under identical settings: $N = 10^4$ trajectories, $10^3$ cubic voxels, and time windows of length $\Delta T = T_0$. Fig.~\ref{fig:K_vs_h_stokes} shows the $\mathcal{K}$--$\mathcal{H}$ relation for inertial particles at $\St=8.89$ (small red circles) and, for comparison, at $\St=0.76$ (large blue circles). The solid black line indicates the linear fit to the $\St=8.89$ dataset. The scatter about the fit is essentially unchanged at higher $\St$, and the fitted slopes are indistinguishable within uncertainty. This indicates that, within the range of Stokes numbers explored here, increasing particle inertia does not measurably alter the  $\mathcal{K}$--$\mathcal{H}$ relation, consistent with trajectory entanglement being controlled mainly by the large-scale structures of the flow.

\subsection{Lagrangian linking in non-turbulent flows}
\label{sec:S3_b}

Having established in the preceding section that the $\mathcal{K}$--$\mathcal{H}$ proportionality holds both locally in space and in time in turbulent flows, we now turn to its limits of validity. Does this proportionality also persist in laminar flows with helicity, or does it require the multi-scale dynamics of turbulence as indicated by the arguments in Section~\ref{sec:S2_a}? Exploring such cases allows us to assess whether the linear correlation should be regarded as a universal property of helical flows, or rather as a distinctive signature of turbulence.

\begin{figure}[t!]
\includegraphics[width=0.32\linewidth, trim={3.5cm 1cm 1cm 1.5cm},clip]{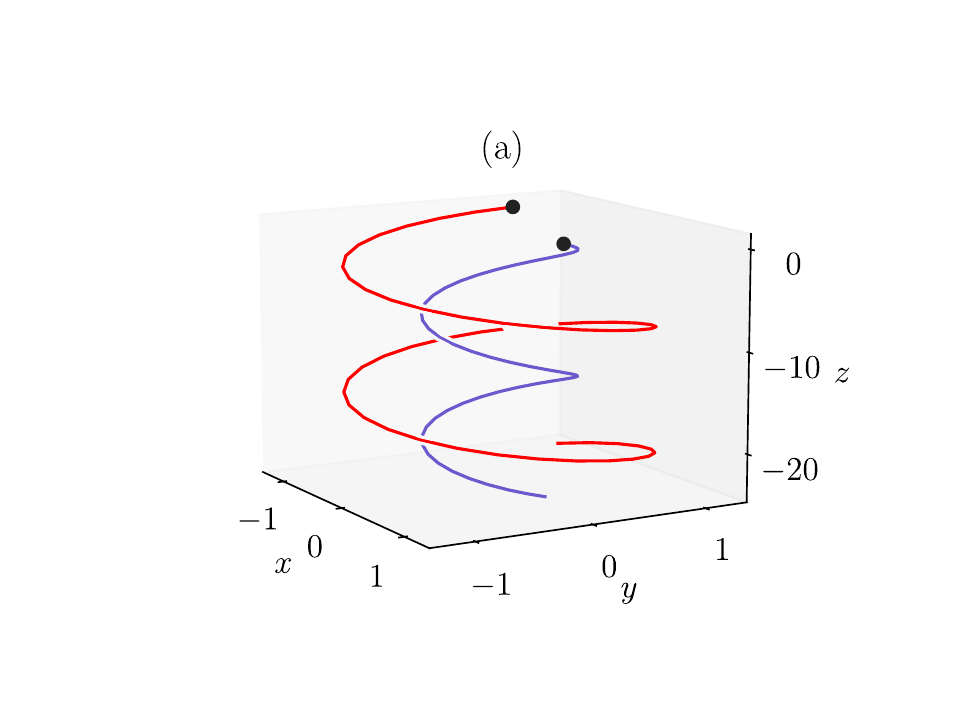}\hspace{1mm}%
\includegraphics[width=0.32\linewidth]{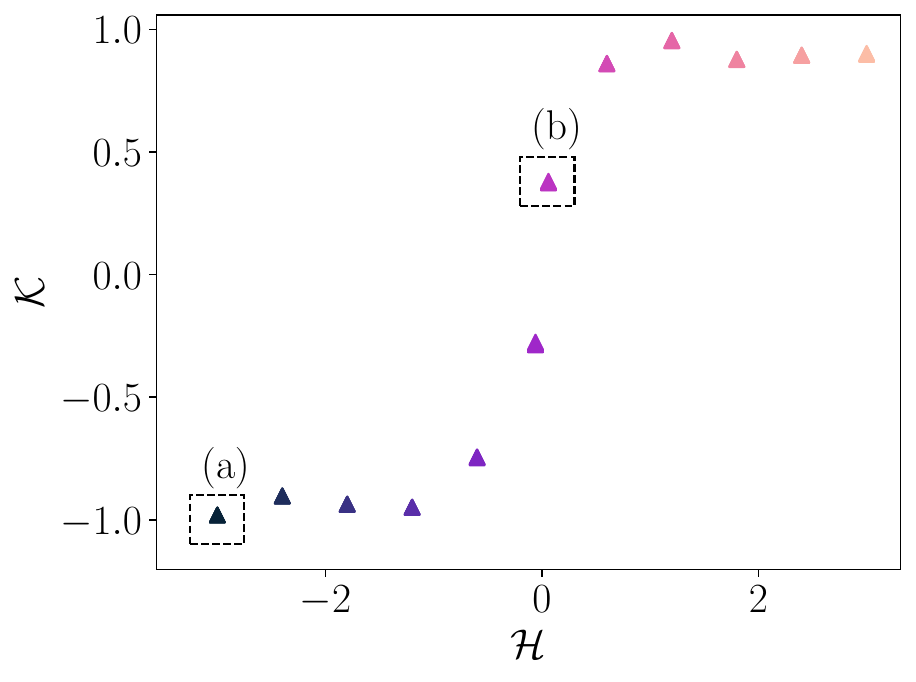}\hspace{5mm}%
\includegraphics[width=0.32\linewidth, trim={3.5cm 1cm 1cm 1.5cm},clip]{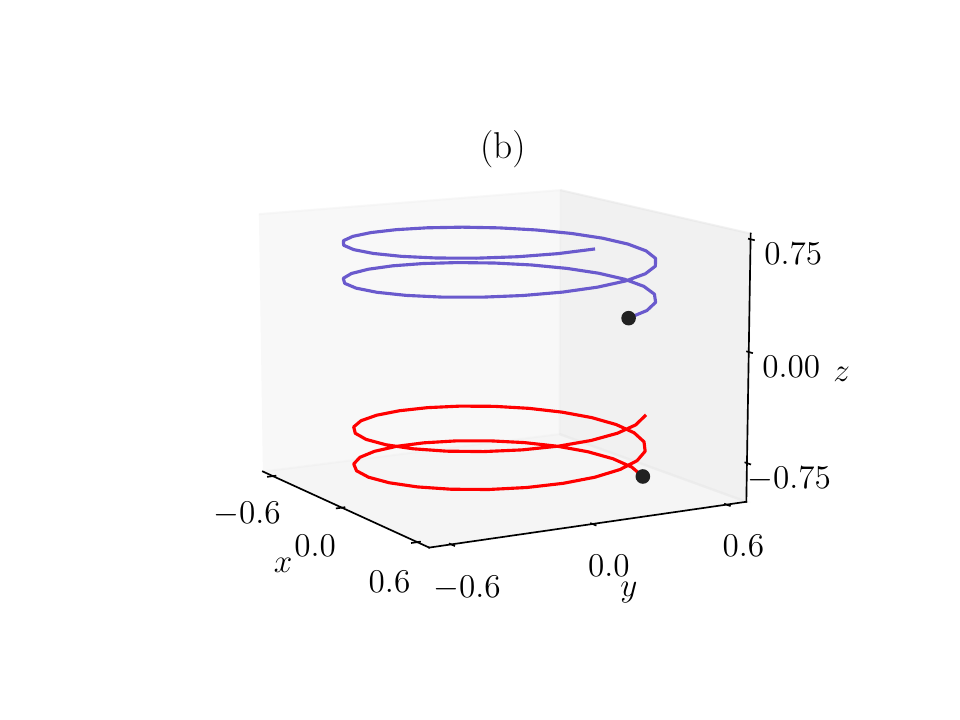}
\caption{Mean linking number $\mathcal{K}$ for ensembles of helical trajectories in the synthetic helix flow, plotted as a function of helicity $\mathcal{H}$. In this configuration, varying the translational velocity directly controls $\mathcal{H}$, so a sweep in $\vec{U}$ corresponds to a continuous change of helicity from negative to positive values.
 The side panels depict two representative cases highlighted by dashed boxes in the central panel: (a) negative helicity, where trajectories are genuinely linked, and (b) positive helicity close to zero, where trajectories are not entangled but their projection generate numerous apparent crossings. The contrast illustrates how projection effects can attenuate the measured value of $\mathcal{K}$ despite de absence of true three-dimensional linking.}
\label{fig:helices_uniform}
\end{figure}

To examine this question, we constructed a synthetic laminar configuration with spatially uniform helicity. This steady flow is prescribed analytically as a homogeneous translation superimposed with solid-body rotation,
\begin{equation}
  \mathbf{u}(\mathbf{r}) = \mathbf{U} + \frac{1}{2} \, \boldsymbol{\Omega} \times \mathbf{r},
\end{equation}
where $\vec{U}$ and $\frac{1}{2} \vec{\Omega}$ denote constant translational and rotational velocity vectors, respectively. In this case, the vorticity is given by $\vec{\omega} = \vec{\Omega}$, and the helicity density reduces to the simple scalar product $H = \vec{U} \cdot \vec{\Omega}$.
Ensembles of 300 tracer trajectories were then evolved in this synthetic flow for different values of $\vec{U}$, thereby sweeping helicity continuously from negative to positive values. The resulting mean linking numbers are shown in the central panel of Fig.~\ref{fig:helices_uniform}.

The results display a distinctly binary response. When tracers evolve in a flow with positive helicity, most projected crossings carry positive sign and the mean linking number approaches $\mathcal{K} \approx +1$. Conversely, in flows with negative helicity, the crossings are predominantly negative and $\mathcal{K} \approx -1$. Once the translational velocity $\vec{U}$ is sufficiently large to generate significant helical motion, $\mathcal{K}$ saturates at those limiting values, and further increases in the magnitude of $\vec{U}$ produce no additional change.

A sharp transition occurs as $\mathcal{H}$ approaches zero. In this regime the flow resembles a planar vortex, and tracer trajectories are confined to nearly two-dimensional orbits with relatively little motion along $\vec{U}$. Genuine entanglement is therefore minimal, and the mean linking number tends to vanish. Yet $\mathcal{K}$ does not collapse exactly to zero, because many apparent crossings arise in projections without true intertwining of the three-dimensional trajectories themselves. This effect is illustrated in the side panels of Fig.~\ref{fig:helices_uniform}: while the left panel shows genuine three-dimensional entanglement, the right panel depicts overlapping rings whose projections generate numerous intersections. In such cases the signed contributions cancel, but the total number of crossings is artificially increased, thereby reducing the average and displaying a slow convergence. As a result, even the vast majority of genuine crossings share the same sign, the mean linking number is attenuated by the abundance of spurious intersections introduced by projection.

The binary outcome of the steady uniform helix flow reflects the fact that, in a time-independent and spatially homogeneous velocity field, trajectory entanglement is dictated solely by geometry. This, in practice, breaks the linear scaling between $\mathcal{K}$ and $\mathcal{H}$ needed for the tomographic method. To probe the influence of temporal variability, we next considered a synthetical helical flow in which helicity oscillates periodically. The velocity field is prescribed as
\begin{equation}
  \vec{u}(\vec{r}, t) = \vec{U} \cos\left(2 \pi f_{\text{mod}} t \right) + \frac{1}{2} \, \vec{\Omega} \times \vec{r},
\end{equation}
so that $\vec{\omega} = \vec{\Omega}$ and the helicity is $H(t) = \vec{U} \cdot \vec{\Omega} \, \cos(2 \pi f_\textrm{mod} t)$, where $f_\textrm{mod}$ is the modulation frequency. Setting $f_\textrm{mod} = 0$ recovers the steady uniform-helicity case described above. Following the same procedure as for that case, we integrated ensembles of 300 tracers in this modulated flow for 200 modulation periods $(T_\textrm{mod} = 1/f_\textrm{mod})$. The resulting
data were analyzed to assess how $\mathcal{K}$ correlates with $\mathcal{H}$ under different coarse-graining windows $\Delta T$.

\begin{figure}[t!]
    \centering
    \includegraphics[width=0.95\linewidth]{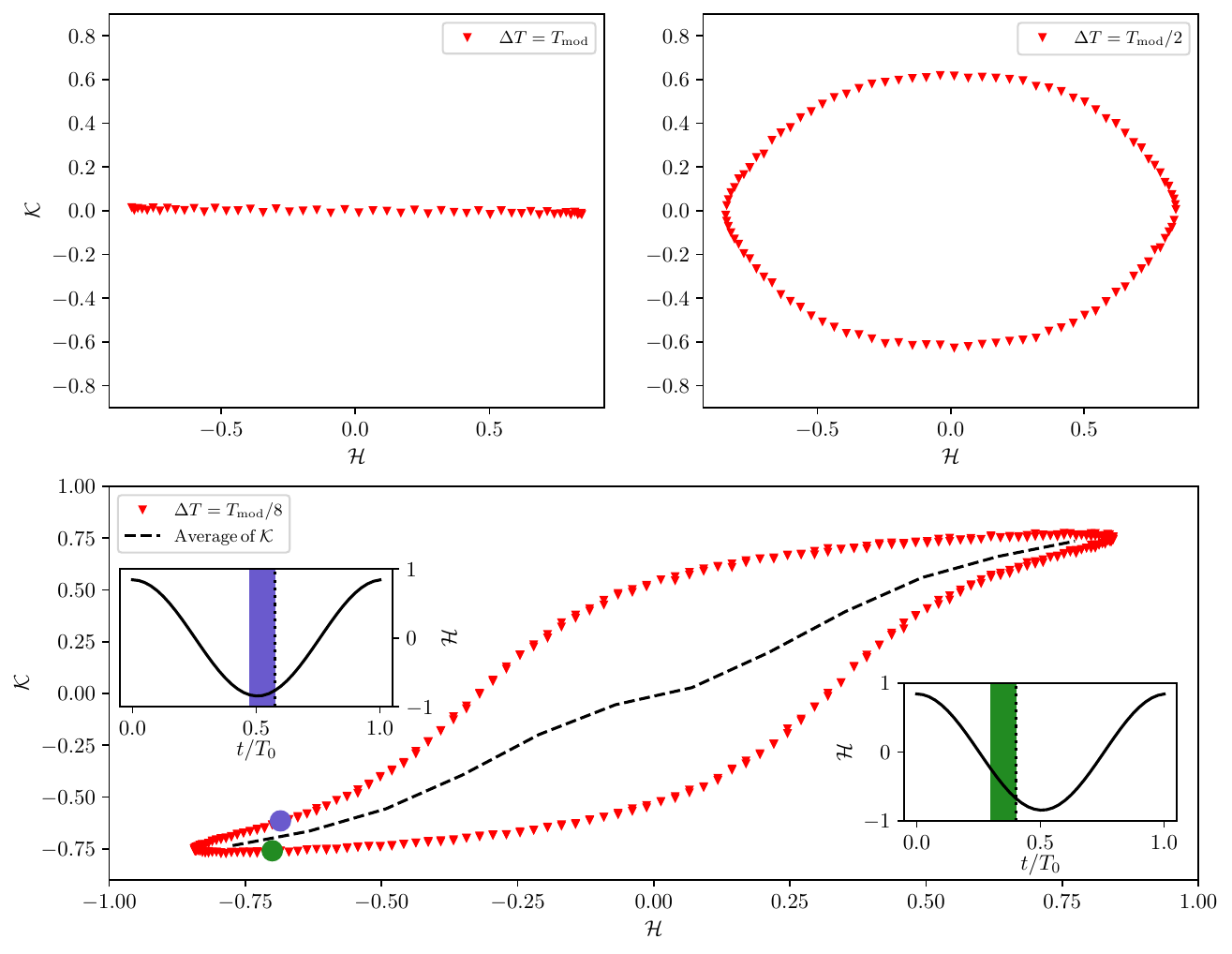}
    \caption{Mean linking number as a function of dimensionless helicity, for the time-modulated helical flow, computed with different time-averaging windows. Top row (left to right): (a) $\Delta T = T_\textrm{mod}$, (b) $\Delta T = T_\textrm{mod}/2$; bottom: (c) $\Delta T = T_\textrm{mod}/8$. For $\Delta T = T_\textrm{mod}/2$, a hysteresis-like behavior emerges in which a given helicity value is compatible with two distinct branches of $\mathcal{K}$. This splitting arises because identical $\mathcal{H}$ values can occur during phases of increasing or decreasing helicity (left and right insets in the bottom panel, respectively), corresponding to different trajectory histories: particles moving upward in the first case, and downward in the second.}
    \label{fig:helices}
\end{figure}

We varied the coarse-graining window $\Delta T$ from one modulation period down to $T_\textrm{mod}/8$, introducing a small offset of $T_0/200$ to sample different phases of the modulation across the 200 periods. As expected, the resulting behavior, shown in Fig.~\ref{fig:helices}, confirms that no correlation emerges when $\Delta T \approx T_\textrm{mod}$: over a full cycle the helicity alternates sign, and the signed trajectory crossings average out, yielding a net mean linking number close to zero.

A more interesting regime is observed when the window is reduced to $\Delta T \simeq T_{\text{mod}}/2$, in which the distribution of $\mathcal{K}$ splits into two distinct branches, as shown in Fig.~\ref{fig:helices}(b). This bifurcation reflects the fact that the same helicity value can be sampled in two different dynamical contexts: during the half-cycle when helicity is positive, or during the half-cycle when it is negative. Within each of these intervals, particles may be moving upward or downward, and their deceleration influences the sign of the counted crossings. The coexistence of these opposing contributions produces two separate behaviors in $\mathcal{K}$, so that the measured linking number depends not only on the instantaneous value of $\mathcal{H}$ but also on the phase of the helicity modulation.

When the averaging window is further shortened to $\Delta T \approx T_\textrm{mod}/8$, as the window is more localized in time, the correlation between signed crossings and helicity increases markedly. This is expected, since within such a short interval the helicity can be regarded as nearly constant, so that trajectory crossings are accumulated under a well-defined sign. As $\Delta T$ decreases further, the analysis approaches the idealized limit of constant helicity within the window, yielding closer agreement between $\mathcal{K}$ and $\mathcal{H}$. In practice, however, excessively small windows are not feasible: the number of detected crossings diminishes rapidly, undermining the statistical reliability of the linking estimate.

This analysis helps explain why a linear relation between $\mathcal{K}$ and $\mathcal{H}$ arises in turbulence but not in laminar or time-modulated flows. In turbulent, statistically stationary conditions, large-scale alignment between velocity and vorticity sets a persistent sign bias in helicity, while fluctuations perturb particle motion and open up many possible linking histories through which signed crossings accumulate. Each realization can be viewed as a microstate of trajectory entanglement: although individual microstates may deviate from the mean, their collective average recovers a robust proportionality between trajectory linking and helicity. The existence of many admissible braiding histories means that several microstates are compatible with the same coarse-grained helicity. This intrinsic degeneracy provides an explanation for both the persistence of the $\mathcal{K}$--$\mathcal{H}$ correlation and the finite spread of $\mathcal{K}$ observed at fixed $\mathcal{H}$. This dispersion is not sampling noise; it reflects the diversity of admissible linking configurations. While increasing particle number, observation time, or tailoring voxel geometry does reduce statistic uncertainty, an intrinsic floor remains, set by the multiplicity of flow microstates consistent with a given helicity level.

Lastly, we explore whether this correlation is a specific feature of turbulence or rather a general property of flows with broad spectral content. To this end, we constructed a synthetic flow as an  analytical superposition of helical waves,
$\mathbf{h}^\pm(\mathbf{k}) = \mathbf{e}(\mathbf{k}) \times (\mathbf{k}/k) \pm i \mathbf{e}(\mathbf{k})$, which are eigenfunctions of the curl operator and form a complete base for incompressible velocity fields \cite{Craya1957, Herring1974}. Here, $\vec{e}(\vec{k})$ is a unit vector orthogonal to $\vec{k}$. In this representation helicity is simply expressed as
\begin{equation}
    H(t) = \sum_\mathbf{k} \, k \, \bigg( |\Tilde{\mathbf{u}}_\mathbf{k}^+(t)|^2 - |\Tilde{\mathbf{u}}_\mathbf{k}^-(t)|^2 \bigg),
\end{equation}
where $\Tilde{\mathbf{u}}_\mathbf{k}^\pm$ denote the velocity-field projections onto the Craya-Herring helical basis for wavenumber $\vec{k}$. This helical–Fourier decomposition separates contributions of opposite mirror symmetry and has been employed in several spectral studies of incompressible turbulent flows (see, e.g., \cite{Alexakis20181} and references therein).

 \begin{figure}[t!]
    \begin{minipage}[t]{0.48\textwidth}
      \raggedright
      (a)\\
      \includegraphics[width=\linewidth]{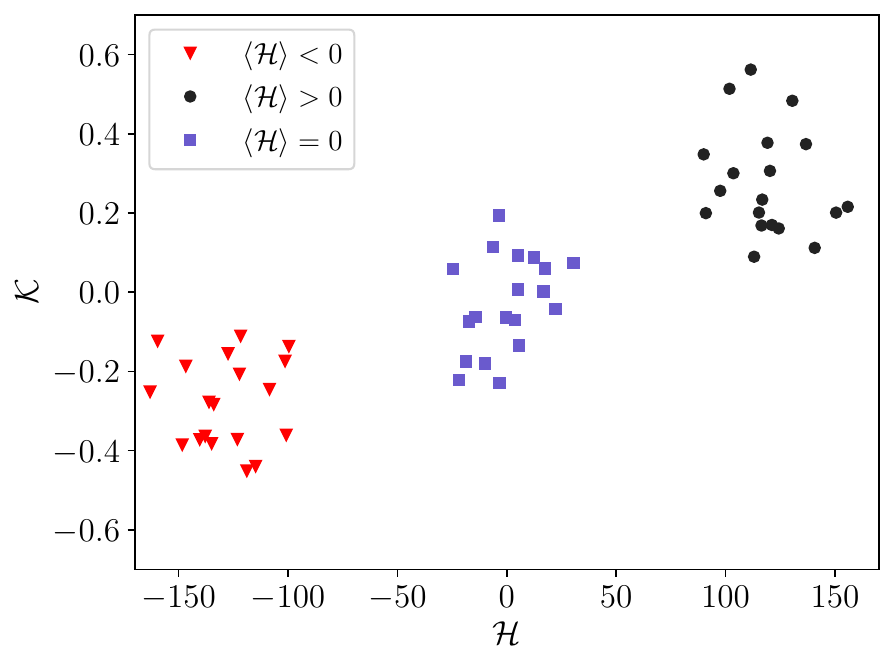}\hspace{0.5cm}
    \end{minipage}\hfill
    \begin{minipage}[t]{0.48\textwidth}
      \raggedright
      (b)\\
      \includegraphics[width=\linewidth]{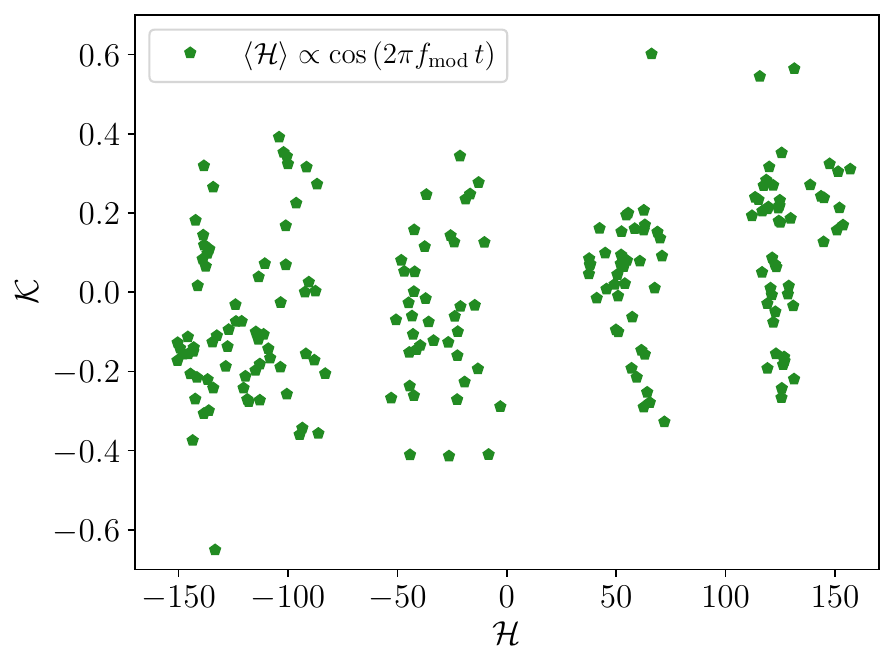}
    \end{minipage}
    \caption{Mean linking number as a function of the dimensionless helicity in synthetic flows constructed from helical Fourier modes. (a) Tracer ensembles in flows with positive (black circles), negative (red triangles), and vanishing mean helicity (blue squares) show a linear correlation between $\mathcal{K}$ and $\mathcal{H}$. (b) Case with time-modulated helicity, where the correlation is lost, highlighting the role of statistical stationarity in sustaining the relation.}
    \label{fig:noise}
\end{figure}

We generated flows by combining $N_k = 100$ such modes, with amplitudes modulated by a stochastic process $\eta(t)$ with unit mean and whose power spectrum follows a $f^{-1}$ scaling over the frequency range $[f_m, f_\eta]$. The resulting field is given by
\begin{equation}
    \mathbf{u}(\mathbf{r}, t) = \eta(t)\sum_\mathbf{k} \bigg[ \mathbf{e}(\mathbf{k}) \times (\mathbf{k}/k) + i \beta_k \mathbf{e}(\mathbf{k}) \bigg] e^{i \mathbf{k} \cdot \mathbf{r}-2\pi f t},
\end{equation}
where a linear dispersion relation $f \propto |\vec{k}|$ is assumed. Three cases were analyzed: $\beta_\mathbf{k} = 1$ (positive helicity), $\beta_\mathbf{k} = -1$ (negative helicity), and $\beta_\mathbf{k} = (-1)^k$ (zero mean helicity).

The results, presented in Fig.~\ref{fig:noise}(a), show that a correlation emerges between $\mathcal{K}$ and $\mathcal{H}$ across the three cases when using a time window of $\Delta T = 1/f_m$. It follows that the proportionality between the trajectory-linking number and dimensionless helicity does not rely exclusively on turbulent nonlinear interactions, but can also arise in flows with broad spectral content. In turbulence, what makes this relation distinctive is that the continual exploration of many such configurations consolidates the proportionality and turns it into a robust statistical signature of the flow. This flow is designed to have high helicity but low velocity; as a result, when $H$ is expressed in units of $U^{-2} \, L^{-2}$, it appears larger than in the previous cases where the RMS vorticity, $\Omega$, has the same order than $U \, L^{-1}$.

Motivated by the central role of statistical stationarity identified above, we finally consider a spectrally rich yet non-stationary counterpart built from the same helical basis by modulating the mode polarizations in time. Accordingly,
\begin{equation}
    \mathbf{u}(\mathbf{r}, t) = \sum_\mathbf{k} \left[ \cos(2\pi f_{\text{mod}} \, t) \, \mathbf{e}(\mathbf{k}) \times (\mathbf{k}/k) + i \sin(2\pi f_{\text{mod}} \, t) \, \mathbf{e}(\mathbf{k}) \right] e^{i \mathbf{k} \cdot \mathbf{r}-2\pi f t}.
\end{equation}
In this unsteady regime, the correlation between $\mathcal{K}$ and $\mathcal{H}$ disappears, as shown in Fig.~\ref{fig:noise}(b). The mechanism is straightforward: the polarization cycle imposes periodic sign reversals of helicity, so over time-windows comparable to the modulation period the signed crossings accrued by tracers cancel, and linking no longer tracks helicity. In principle, the proportionality could be recovered in the short window limit ($\Delta T \ll f_\textrm{mod}^{-1}$) given sufficiently large ensembles; in practice, for realistic window lengths and sample sizes, the modulation erases the correlation. Taken together with the stationary helical-mode case analyzed previously, this isolates stationarity over a reasonable time window prescribed by the statistical needs to determine $\mathcal{K}$ ---rather than spectral breadth--- as the condition under which the mean linking number $\mathcal{K}$ serves as a reliable proxy for helicity.

\section{Conclusions}
\label{sec:S4}

Previous work established empirically that the mean helicity of turbulent flows correlates linearly with the global average linking number of tracer trajectories, unveiling a topological signature of turbulence. In the present study we have extended this connection in several fundamental directions. We demonstrated that the entanglement of particle trajectories, quantified through the average signed crossing number $\mathcal{K}$, provides a robust proxy for helicity even locally in space and time. Based on the topological interpretation of helicity, we derived an analytical rationale showing that the ensemble-averaged linking rate is proportional to the coarse-grained helicity density, with a proportionality involving correlation lengths and effective timescales of the flow. This establishes a direct bridge between a topological observable accessible from Lagrangian data and a fundamental dynamical invariant of turbulence, enabling helicity tomography without requiring velocity gradients reconstruction.

From Lagrangian trajectory data stemming from both numerical Taylor-Green and laboratory von K\'arm\'an turbulent flows, we obtained the first trhee-dimensional spatial reconstructions of helicity fields. Spatial maps of $\mathcal{K}$ recover the main chiral structures of both flows, while time series of $\mathcal{K}$ trace the evolution of the domain-averaged helicity. The correlation remains essentially unaffected by finite particle inertia (as observed numerically with particles of Stokes number 0.76 and 8.89), consistent with helicity being dominated by large-scale structures, and proves robust against changes in voxel geometry. These results establish $\mathcal{K}$ as a quantitative probe of helicity in current experimentally attainable conditions (number of trajectories and observation window length) where conventional Eulerian approaches are unfeasible, and show that Lagrangian trajectories alone contain sufficient information to reconstruct the spatial structure of helicity in turbulent flows.

Further analyzing non-turbulent flows, we delineated the regimes where proportionality between $\mathcal{K}$ and $\mathcal{H}$ holds. In laminar flows with uniform helicity, $\mathcal{K}$ saturates to a binary response, reflecting purely geometric entanglement. In synthetic time-modulated helical flows or helical-wave fields with non-stationary polarization, the correlation disappears or becomes hysteretic, with $\mathcal{K}$ depending strongly on sampling windows and trajectory history. These results indicate that the linear relation between trajectory linking and helicity is not a general property of all helical flows, but rather emerges robustly in statistically stationary turbulent regimes.

In turbulent flows, our results show that the $\mathcal{K}$--$\mathcal{H}$ relation retains a finite residual spread, which does not vanish even with a large amount of statistics. This variance reflects the multiplicity of admissible trajectory entanglement histories compatible with a given helicity level, and is therefore intrinsic rather than due to sampling noise. In this sense, helicity constraints the ensemble average but not the detailed realization: many distinct microstates of trajectory linking are consistent with the same coarse-grained value of $\mathcal{H}$. Beyond this intrinsic spread, dispersion scales as expected with particle number, observation time, and voxel geometry, providing operation guidelines for experiments. Thus, the observed scattering itself carries physical meaning, hinting at the degeneracy of Lagrangian pathlines in turbulent helicity transport.

This study opens several directions for future work. The ability to reconstruct helicity fields directly from trajectories suggests immediate applications in laboratory and geophysical flows where velocity gradients are inaccessible. A promising perspective is to couple $\mathcal{K}$-based tomography with event-detection techniques (such as Lagrangian coherent structure diagnostics) to isolate how specific topological changes contribute to helicity production and destruction across scales. 

\section{Acknowledgments}

The authors acknowledge financial support from UBACyT Grant No. 20020220300122BA, and from Proyecto REMATE of the Redes Federales de Alto Impacto, Argentina. PDM acknowledges a fruitful conversation with Darryl Holm.

\section{Data availability}
\label{sec:data}

The code developed in this work (LaLiTo: Lagrangian Linking Tomography) is openly available at Zenodo \cite{NosedaZenodo2025}. An example raw dataset of the Taylor-Green flow with $10^3$ particles of $\St = 0.76$, as well as the processed data are available at the same repository.

\appendix

\section{Trajectory crossing rate for long-runs}
\label{app:rationale-time-averages}

The ensemble-averaged formulation presented in \ref{sec:S2_a} establishes the proportionality between the mean trajectory linking rate and the local helicity density. In practice, however, experiments and simulations often yield a single, sufficiently long realization rather than multiple independent runs. In this Section we present the time-average version of the argument, valid under statistical stationarity and ergodicity, particularly suitable for single, long-run datasets.

We begin with Eq.~\eqref{eq:5}, which gives the per-unit-time mean linking number over a finite window of length $\Delta T$ and within a volume $V$:
\begin{equation}
  \frac{\Lk_{V,\Delta T}}{\Delta T} = \frac{1}{4\pi V^2 \Delta T} \int_V  d^3x_1 \int_V  d^3x_2  \int_{-\Delta T}^{\Delta T}  \int_{|\tau|/2}^{\Delta T - |\tau|/2}
  \frac{\vec{\omega}(\vec{x}_1,T-\tau/2) \cdot \vec{u}(\vec{x}_2,T+\tau/2)}  {|\vec{x}_1-\vec{x}_2|} \: dT \: d\tau.
  \label{eq:A1}
\end{equation}
This representation follows from discarding the boundary contribution in Eq.~\eqref{eq:Lk_2terminos}; which is exact in periodic domains, vanishes for no-slip walls, and is negligible under free-slip in the regimes considered.

To isolate the purely temporal content, we introduce the windowed time average over the time $T$,
\begin{equation}
  \hat{h}_{\Delta T}(\vec{x}_1, \vec{x}_2, \tau) = \frac{1}{\Delta T - |\tau |} \int_{|\tau |/2}^{\Delta T- |\tau |/2} \, \vec{\omega}(\vec{x}_1, T - \tau/2) \cdot \vec{u}(\vec{x}_2, T + \tau/2) \, dT.
  \label{eq:A2}
\end{equation}
Substituting this definition into Eq.~\eqref{eq:A1} and performing the $T$-integration yields
\begin{equation}
  \frac{\Lk_{V,\Delta T}}{\Delta T} = \frac{1}{4\pi V^2 \Delta T} \int_V \int_V  \int_{-\Delta T}^{\Delta T}  \frac{\Delta T - |\tau |}{|\vec{x}_1-\vec{x}_2|} \, \hat{h}_{\Delta T}(\vec{x}_1, \vec{x}_2, \tau) \: d\tau \: d^3x_1 \: d^3x_2,
  \label{eq:A3}
\end{equation}
where the factor $(\Delta T- | \tau |)/\Delta T$ is the (triangular) finite-window weight that counts how many central times $T$ admit a given lag $\tau$. In statistically stationary flows with finite helicity decorrelation time $\tau_h$, and for windows $\Delta T$ much larger than $\tau_h$, this weight may be replaced by unity and the $\tau$-integral extended to $(-\infty, \infty)$ with an error of order $\tau_h/\Delta T$:
\begin{equation}
  \frac{\Lk_{V,\Delta T}}{\Delta T} = \frac{1}{4\pi V^2} \int_V \int_V  \int_{-\infty}^{\infty}  \frac{\hat{h}_{\Delta T}(\vec{x}_1, \vec{x}_2, \tau) }{|\vec{x}_1-\vec{x}_2|} \, \: d\tau \: d^3x_1 \: d^3x_2 + \mathcal{O}(\tau_h/\Delta T).
  \label{eq:A4}
\end{equation}
The following step relies on the hypothesis that the dynamics is not only statistically stationary but also ergodic. Under this additional assumption, the long-time windowed average converges to the ensemble two-time correlator,
\begin{equation}
 \hat{h}_{\Delta T}(\vec{x}_1, \vec{x}_2, \tau) \longrightarrow h(\vec{x}_1, \vec{x}_2, \tau) \equiv \langle \vec{\omega}(\vec{x}_1, t) \cdot \vec{u}(\vec{x}_2, t+\tau) \rangle_\textrm{ens},
  \label{eq:A5}
\end{equation}
as $\Delta T \rightarrow \infty$. Inserting this limit into Eq.~\eqref{eq:A4} recovers Eq.~\eqref{eq:Lkbar_punto_pre_tauh} and, using the definition of  $\chi(\vec{x}_1, \vec{x}_2)$ given in the main text, also reproduces Eq.~\eqref{eq:Lkbar_punto_simplificada}.
From this point on, the time- and ensemble-ageraged formulations coincide, so the locality argument developed in Section~\ref{sec:S2_a} holds without modification here. Expanding around the pair separation $\vec{\xi} = \vec{x}_1 - \vec{x}_2$ and exploiting the near-field isotropy gives the same geometric factor, which leads to
\begin{equation}
  \left\langle \frac{\Lk_{V,\Delta T}}{\Delta T} \right\rangle_\textrm{ens} \approx \frac{\ell_h^2}{2 V} \langle \chi(\vec{x},\vec{x}) \rangle_V.
  \label{eq:A6}
\end{equation}
At coincident points the time-integrated correlator separates as $\chi(\vec{x},\vec{x}) = \langle \vec{u} \cdot \vec{\omega} \rangle \int_{-\infty}^\infty C_{\vec{x}} \, d\tau$, which defines an effective helicity correlation time $\tau_h^\textrm{eff}$. Averaging over the volume $V$ then yields the proportionality stated in Eq.~\eqref{eq:10}, i.e.,
\begin{equation}
  \left\langle \frac{\Lk_{V,\Delta T}}{\Delta T} \right\rangle_\textrm{ens} \approx
  \frac{\ell_h^2}{2 V} \, \tau^{\text{eff}}_h \, \langle \vec{u} \cdot \vec{\omega} \rangle_V.
  \label{eq:A7}
\end{equation}
Consequently, in statistically stationary and ergodic regimes a single sufficiently long realization furnishes an estimator of the ensemble linking rate and, through Eq.~\eqref{eq:A7}, of the local mean helicity. This time-average formulation complements the ensemble-based rationale of Section~\ref{sec:S2_a} and is particularly relevant in experimental and single-run numerical settings, where analyses often rely on one long record rather than multiple realizations.

\bibliography{bibliography}

@article{Mininni2011,
    doi = {10.1016/j.parco.2011.05.004},
    url = {http://dx.doi.org/10.1016/j.parco.2011.05.004},
    year = 2011,
    month = {jun},
	pages={316–326},
    publisher = {Elsevier BV},
    volume = {37},
    number = {6–7},
    author = { Pablo D. Mininni and  Duane Rosenberg and  Raghu Reddy and  Annick Pouquet},
    title = {A hybrid MPI–OpenMP scheme for scalable parallel pseudospectral computations for fluid turbulence},
    journal = {Parallel Computing}
}

@article{Rosenberg2020,
    doi = {10.3390/atmos11020178},
    url = {http://dx.doi.org/10.3390/atmos11020178},
    year = 2020,
    month = {feb, pages={178}},
    publisher = {MDPI AG},
    volume = {11},
    number = {2},
    author = { Duane Rosenberg and  Pablo D. Mininni and  Raghu Reddy and  Annick Pouquet},
    title = {GPU Parallelization of a Hybrid Pseudospectral Geophysical Turbulence Framework Using CUDA},
    journal = {Atmosphere}
}

@article{Angriman2020,
	doi = {10.1103/physrevfluids.5.064605},
	url = {https://doi.org/10.1103/physrevfluids.5.064605},
	year = 2020,
	month = {jun},
	publisher = {American Physical Society ({APS})},
	volume = {5},
	number = {6},
	author = {Sof{\'{\i}}a Angriman and Pablo D. Mininni and Pablo J. Cobelli},
	title = {Velocity and acceleration statistics in particle-laden turbulent swirling flows},
	journal = {Physical Review Fluids}
}

@article{Angriman2022,
	doi = {10.1017/jfm.2022.713},
	url = {https://doi.org/10.1017/jfm.2022.713},
	year = 2022,
	month = {sep},
	publisher = {Cambridge University Press ({CUP})},
	volume = {948},
	author = {Sof{\'{\i}}a Angriman and Am{\'{e}}lie Ferran and Florencia Zapata and Pablo J. Cobelli and Martin Obligado and Pablo D. Mininni},
	title = {Clustering in laboratory and numerical turbulent swirling flows},
	journal = {Journal of Fluid Mechanics}
}

@article{Angriman2022_2,
	doi = {10.1103/physrevfluids.7.064603},
	url = {https://doi.org/10.1103/physrevfluids.7.064603},
	year = 2022,
	month = {jun},
	publisher = {American Physical Society ({APS})},
	volume = {7},
	number = {6},
	author = {Sof{\'{\i}}a Angriman and Pablo D. Mininni and Pablo J. Cobelli},
	title = {Multitime structure functions and the Lagrangian scaling of turbulence},
	journal = {Physical Review Fluids}
}

@article{Angriman_2021,
	author = {S. Angriman and P.~J. Cobelli and M. Bourgoin and S.~G. Huisman and R. Volk and P.~D. Mininni},
	journal = {Physical Review Letters},
	month = {dec},
	number = 25,
	pages = 254502,
	title = {Broken Mirror Symmetry of Tracer's Trajectories in Turbulence},
	volume = 127,
	year = 2021,
  url = {https://doi.org/10.1103/PhysRevLett.127.254502}
}

@article{Berning2023,
    author = {Berning, Hanna and Rösgen, Thomas},
    title = "{Suppression of large-scale azimuthal modulations in a von Kármán flow using random forcing}",
    journal = {Physics of Fluids},
    volume = {35},
    number = {7},
    pages = {075151},
    year = {2023},
    month = {07},
    issn = {1070-6631},
    doi = {10.1063/5.0152876},
    url = {https://doi.org/10.1063/5.0152876},
}

@article{Espanol2025,
  title = {Effect of local flow geometry on particle pair dispersion angle},
  author = {Espa\~nol, B. L. and Noseda, M. and Cobelli, P. J. and Mininni, P. D.},
  journal = {Phys. Rev. Fluids},
  volume = {10},
  issue = {4},
  pages = {044501},
  numpages = {21},
  year = {2025},
  month = {Apr},
  publisher = {American Physical Society},
  doi = {10.1103/PhysRevFluids.10.044501},
  url = {https://link.aps.org/doi/10.1103/PhysRevFluids.10.044501}
}

@book{Kauffman2001,
author = {Kauffman, Louis H},
title = {Knots and Physics},
publisher = {World Scientific},
year = 2001,
doi = {10.1142/4256},
edition   = {3rd},
URL = {https://www.worldscientific.com/doi/abs/10.1142/4256},
}

@article{Alexakis20181,
title = {Cascades and transitions in turbulent flows},
journal = {Physics Reports},
volume = {767-769},
pages = {1-101},
year = {2018},
note = {Cascades and transitions in turbulent flows},
issn = {0370-1573},
doi = {https://doi.org/10.1016/j.physrep.2018.08.001},
url = {https://www.sciencedirect.com/science/article/pii/S0370157318301935},
author = {A. Alexakis and L. Biferale}}

@article{Harms2023,
author = {Harms, Tanner D.  and Brunton, Steven L.  and McKeon, Beverley J. },
title = {Lagrangian gradient regression for the detection of coherent structures from sparse trajectory data},
journal = {Royal Society Open Science},
volume = {11},
number = {10},
pages = {240586},
year = {2024},
doi = {10.1098/rsos.240586},
URL = {https://royalsocietypublishing.org/doi/abs/10.1098/rsos.240586}}

@article{Ferraro2024,
    author = {Ferraro, D. and Servidio, S. and Lauria, A. and Gaudio, R.},
    title = {A local measure of the helicity in turbulent flows},
    journal = {Physics of Fluids},
    volume = {36},
    number = {10},
    pages = {105146},
    year = {2024},
    month = {10},
    issn = {1070-6631},
    doi = {10.1063/5.0223162},
    url = {https://doi.org/10.1063/5.0223162},
}

@inbook{Yokoi2023,
author = {Yokoi, Nobumitsu},
publisher = {American Geophysical Union (AGU)},
isbn = {9781119841715},
title = {Transport in Helical Fluid Turbulence},
booktitle = {Helicities in Geophysics, Astrophysics and Beyond},
chapter = {3},
pages = {25-50},
doi = {https://doi.org/10.1002/9781119841715.ch3},
url = {https://agupubs.onlinelibrary.wiley.com/doi/abs/10.1002/9781119841715.ch3},
year = {2023}}

@article{Moffatt1969,
title={The degree of knottedness of tangled vortex lines},
volume={35},
DOI={10.1017/S0022112069000991},
number={1},
journal={Journal of Fluid Mechanics},
author={Moffatt, H. K.},
year={1969},
pages={117–129}}

@article{
Scheeler2014,
author = {Martin W. Scheeler  and Dustin Kleckner  and Davide Proment  and Gordon L. Kindlmann  and William T. M. Irvine },
title = {Helicity conservation by flow across scales in reconnecting vortex links and knots},
journal = {Proceedings of the National Academy of Sciences},
volume = {111},
number = {43},
pages = {15350-15355},
year = {2014},
doi = {10.1073/pnas.1407232111},
URL = {https://www.pnas.org/doi/abs/10.1073/pnas.1407232111},
}

@article{Wallace1992,
    author = {Wallace, James M. and Balint, Jean‐Louis and Ong, Lawrence},
    title = {An experimental study of helicity density in turbulent flows},
    journal = {Physics of Fluids A: Fluid Dynamics},
    volume = {4},
    number = {9},
    pages = {2013-2026},
    year = {1992},
    month = {09},
    issn = {0899-8213},
    doi = {10.1063/1.858371},
    url = {https://doi.org/10.1063/1.858371},
}

@article{
Scheeler2017,
author = {Martin W. Scheeler  and Wim M. van Rees  and Hridesh Kedia  and Dustin Kleckner  and William T. M. Irvine },
title = {Complete measurement of helicity and its dynamics in vortex tubes},
journal = {Science},
volume = {357},
number = {6350},
pages = {487-491},
year = {2017},
doi = {10.1126/science.aam6897},
URL = {https://www.science.org/doi/abs/10.1126/science.aam6897},
}

@ARTICLE{Schanz2016,
  title     = "{Shake-The-Box}: Lagrangian particle tracking at high particle
               image densities",
  author    = "Schanz, Daniel and Gesemann, Sebastian and Schr{\"o}der, Andreas",
  journal   = "Experiments in Fluids",
  publisher = "Springer Science and Business Media LLC",
  volume    =  57,
  number    =  5,
  month     =  may,
  year      =  2016,
  url = {https://doi.org/10.1007/s00348-016-2157-1}
}

@article{Schroder2023,
   author = "Schröder, Andreas and Schanz, Daniel",
   title = "3D Lagrangian Particle Tracking in Fluid Mechanics", 
   journal= "Annual Review of Fluid Mechanics",
   year = "2023",
   volume = "55",
   number = "Volume 55, 2023",
   pages = "511-540",
   doi = "https://doi.org/10.1146/annurev-fluid-031822-041721",
   url = "https://www.annualreviews.org/content/journals/10.1146/annurev-fluid-031822-041721",
   publisher = "Annual Reviews",
   issn = "1545-4479",
   type = "Journal Article",
  }

@article{Waleffe1992,
  author       = {Waleffe, F.},
  title        = {The nature of triad interactions in homogeneous turbulence},
  journal      = {Physics of Fluids A},
  year         = {1992},
  volume       = {4},
  number       = {2},
  pages        = {350--363},
  doi          = {10.1063/1.858309},
}

@article{Biferale2012,
  author       = {Biferale, Luca and Musacchio, Stefano and Toschi, Federico},
  title        = {Inverse energy cascade in three-dimensional isotropic turbulence},
  journal      = {Physical Review Letters},
  year         = {2012},
  volume       = {108},
  number       = {16},
  pages        = {164501},
  doi          = {10.1103/PhysRevLett.108.164501},
}

@article{Chen2003,
  author       = {Chen, Qiaoning and Chen, Shiyi and Eyink, Gregory L.},
  title        = {The joint cascade of energy and helicity in three-dimensional turbulence},
  journal      = {Physics of Fluids},
  year         = {2003},
  volume       = {15},
  number       = {2},
  pages        = {361--374},
  doi          = {10.1063/1.1533070},
}

@article{Padhye1996b,
  title = {Relabeling Symmetries in Hydrodynamics and Magnetohydrodynamics},
  journal = {Plasma Physics Reports},
  author = {Padhye, N. and Morrison, P.J.},
  year = {1996},
  volume = {22},
  number = {10},
  pages = {869-877},
  doi = {10.2172/226406},
}

@article{Morrison1998,
  title = {Hamiltonian Description of the Ideal Fluid},
  author = {Morrison, P J},
  year = {1998},
  journal = {Rev. Mod. Phys.},
  volume = {70},
  number = {2},
  langid = {english},
  url = {https://doi.org/10.1103/RevModPhys.70.467}
}

@article{Salmon1988,
   author = "Salmon, R",
   title = "Hamiltonian Fluid Mechanics",
   journal= "Annual Review of Fluid Mechanics",
   year = "1988",
   volume = "20",
   pages = "225-256",
   doi = "https://doi.org/10.1146/annurev.fl.20.010188.001301",
   url = "https://www.annualreviews.org/content/journals/10.1146/annurev.fl.20.010188.001301",
   publisher = "Annual Reviews",
   issn = "1545-4479",
   type = "Journal Article",
  }

@BOOK{DeBerg2000,
  title     = "Computational geometry",
  author    = "de Berg, Mark and van Kreveld, Marc and Overmars, Mark and Schwarzkopf, Otfried",
  publisher = "Springer",
  edition   =  2,
  year      =  2000,
  address   = "Berlin, Germany",
}

@ARTICLE{Bentley1979,
  author={Bentley and Ottmann},
  journal={IEEE Transactions on Computers},
  title={Algorithms for Reporting and Counting Geometric Intersections},
  year={1979},
  volume={C-28},
  number={9},
  pages={643-647},
  doi={10.1109/TC.1979.1675432}}

@article{Moffatt1992,
   author = "Moffatt, H. K. and Tsinober, A.",
   title = "Helicity in Laminar and Turbulent Flow",
   journal= "Annual Review of Fluid Mechanics",
   year = "1992",
   volume = "24",
   pages = "281-312",
   doi = "https://doi.org/10.1146/annurev.fl.24.010192.001433",
   url = "https://www.annualreviews.org/content/journals/10.1146/annurev.fl.24.010192.001433",
   publisher = "Annual Reviews",
   issn = "1545-4479",
  }

@BOOK{Raffel2007,
  title     = "Particle Image Velocimetry",
  author    = "Raffel, Markus and Willert, C E and Wereley, Steven T and Kompenhans, Jurgen",
  publisher = "Springer",
  edition   =  2,
  year      =  2007,
  address   = "Berlin, Germany",
}

@ARTICLE{Taylor1937,
  title     = "Mechanism of the production of small eddies from large ones",
  author    = "Taylor, G I and Green, A E",
  journal   = "Proc. R. Soc. Lond.",
  publisher = "The Royal Society",
  volume    =  158,
  number    =  895,
  pages     = "499--521",
  year      =  1937,
  url= {https://doi.org/10.1098/rspa.1937.0036}
}

@ARTICLE{Maxey1983,
  title     = "Equation of motion for a small rigid sphere in a nonuniform flow",
  author    = "Maxey, Martin R and Riley, James J",
  journal   = "Phys. Fluids",
  publisher = "AIP Publishing",
  volume    =  26,
  number    =  4,
  pages     = "883--889",
  year      =  1983,
  url       = {https://doi.org/10.1063/1.864230}
}

@article{Gatignol1983,
  author = {Gatignol, R.},
  journal = {J. Mec. Theor. Appl.},
  pages = {143--160},
  title = {The {F}ax\'{e}n Formul{\ae} for a Rigid Particle in an Unsteady Non-uniform {S}tokes Flow},
  volume = 1,
  year = 1983
}

@article{Auton1988,
title={The force exerted on a body in inviscid unsteady non-uniform rotational flow},
volume={197},
DOI={10.1017/S0022112088003246},
journal={Journal of Fluid Mechanics},
author={Auton, T. R. and Hunt, J. C. R. and Prud’Homme, M.},
year={1988},
pages={241–257}
}

@article{Boussinesq1885,
author="Boussinesq, J.",
title="Sur la resistance qu'oppose un fluide indefini en repos, sans pesanteur, au mouvement varie d'une sphere solide qu'il mouille sur toute sa surface, quand les vitesses restent bien continues et assez faibles pour que leurs carres et produits soient negligiables",
journal="C. R. Acad. Sc. Paris",
year="1885",
volume="100",
pages="935-937",
URL="https://cir.nii.ac.jp/crid/1571135650072086912"
}

@book{Basset1888,
  title={A Treatise on Hydrodynamics: With Numerous Examples},
  author={Basset, A.B.},
  volume={1},
  lccn={02022039},
  series={A Treatise on Hydrodynamics},
  year={1888},
  publisher={Deighton, Bell and Company}
}

@Inbook{Happel1983,
author="Happel, John and Brenner, Howard",
title="Wall Effects on the Motion of a Single Particle",
bookTitle="Low Reynolds number hydrodynamics: with special applications to particulate media",
year="1983",
publisher="Springer Netherlands",
address="Dordrecht",
pages="286--357",
isbn="978-94-009-8352-6",
doi="10.1007/978-94-009-8352-6_7",
url="https://doi.org/10.1007/978-94-009-8352-6_7"
}

@BOOK{Adrian2010,
  title     = "Particle image velocimetry series",
  author    = "Adrian, Ronald J and Westerweel, Jerry",
  publisher = "Cambridge University Press",
  year      =  2010,
  address   = "Cambridge, England"
}

@article{BergerField1984, title={The topological properties of magnetic helicity}, volume={147}, DOI={10.1017/S0022112084002019}, journal={Journal of Fluid Mechanics}, author={Berger, Mitchell A. and Field, George B.}, year={1984}, pages={133–148}}

@article{Teitelbaum2009,
  title = {Effect of Helicity and Rotation on the Free Decay of Turbulent Flows},
  author = {Teitelbaum, T. and Mininni, P. D.},
  journal = {Phys. Rev. Lett.},
  volume = {103},
  issue = {1},
  pages = {014501},
  numpages = {4},
  year = {2009},
  month = {Jun},
  publisher = {American Physical Society},
  doi = {10.1103/PhysRevLett.103.014501},
  url = {https://link.aps.org/doi/10.1103/PhysRevLett.103.014501}
}

@article{Pouquet_Patterson_1978, title={Numerical simulation of helical magnetohydrodynamic turbulence}, volume={85}, DOI={10.1017/S0022112078000658}, number={2}, journal={Journal of Fluid Mechanics}, author={Pouquet, A. and Patterson, G. S.}, year={1978}, pages={305–323}}

@article{Volk2008,
doi = {10.1209/0295-5075/81/34002},
url = {https://dx.doi.org/10.1209/0295-5075/81/34002},
year = {2007},
month = {dec},
publisher = {},
volume = {81},
number = {3},
pages = {34002},
author = {Volk, R. and Mordant, N. and Verhille, G. and Pinton, J.-F.},
title = {Laser Doppler measurement of inertial particle and bubble accelerations in turbulence},
journal = {Europhysics Letters}
}

@article{Mordant2002,
    author = {Mordant, N. and Metz, P. and Pinton, J.‐F. and Michel, O.},
    title = {Lagrangian measurement in fully developed turbulence},
    journal = {AIP Conference Proceedings},
    volume = {622},
    number = {1},
    pages = {343-352},
    year = {2002},
    month = {07},
    issn = {0094-243X},
    doi = {10.1063/1.1487552},
    url = {https://doi.org/10.1063/1.1487552},
}

@article{Padhye1996,
title = {Fluid element relabeling symmetry},
journal = {Physics Letters A},
volume = {219},
number = {5},
pages = {287-292},
year = {1996},
issn = {0375-9601},
doi = {https://doi.org/10.1016/0375-9601(96)00472-0},
url = {https://www.sciencedirect.com/science/article/pii/0375960196004720},
author = {Nikhil Padhye and P.J. Morrison}
}

@article{Holm1999,
title = {Fluctuation effects on 3D Lagrangian mean and Eulerian mean fluid motion},
journal = {Physica D: Nonlinear Phenomena},
volume = {133},
number = {1},
pages = {215-269},
year = {1999},
issn = {0167-2789},
doi = {https://doi.org/10.1016/S0167-2789(99)00093-7},
url = {https://www.sciencedirect.com/science/article/pii/S0167278999000937},
author = {Darryl D. Holm}
}

@phdthesis{Craya1957,
  author    = {Antoine Craya},
  title     = {Contribution à l'analyse de la turbulence associée à des vitesses moyennes},
  school    = {Université de Grenoble},
  year      = 1957,
  type      = {Thèse de doctorat},
  language  = {French},
    URL = {https://theses.hal.science/tel-00684659},

}

@article{Herring1974, title={Decay of two-dimensional homogeneous turbulence}, volume={66}, DOI={10.1017/S0022112074000280}, number={3}, journal={Journal of Fluid Mechanics}, author={Herring, J. R. and Orszag, S. A. and Kraichnan, R. H. and Fox, D. G.}, year={1974}, pages={417–444}}

@article{Wang1993, title={Settling velocity and concentration distribution of heavy particles in homogeneous isotropic turbulence}, volume={256}, DOI={10.1017/S0022112093002708}, journal={Journal of Fluid Mechanics}, author={Wang, Lian-Ping and Maxey, Martin R.}, year={1993}, pages={27–68}}

@article{Moreau1961,
  author    = {Moreau, Jean~Jacques},
  title     = {Constantes d’un îlot tourbillonnaire en fluide parfait barotrope},
  journal   = {Comptes rendus hebdomadaires des séances de l’Académie des sciences},
  volume    = {252},
  pages     = {2810--2812},
  PUBLISHER = {{Gauthier-Villars}},
  year      = {1961},
  language  = {French},
  URL = {https://hal.science/hal-01865239},
}

@software{NosedaZenodo2025,
  author       = {Noseda, M. and Cobelli, P.~J.},
  title        = {LaLiTo: Lagrangian Linking Tomography — Signed crossing density mapping from 3D Lagrangian trajectories},
  year         = 2025,
  publisher    = {Zenodo},
  doi          = {10.5281/zenodo.17070146},
  url          = {https://doi.org/10.5281/zenodo.17070146}
}

@incollection{Gauss1833,
  author    = {Gauss, Carl Friedrich},
  title     = {Integral formula for linking number},
  booktitle = {Zur mathematischen Theorie der electrodynamischen Wirkungen},
  editor    = {K{\"o}niglichen Gesellschaft der Wissenschaften zu G{\"o}ttingen},
  publisher = {Springer Berlin Heidelberg},
  year      = {1833},
  volume    = {5},
  pages     = {605},
  address   = {Berlin, Heidelberg},
}

@article{Berger2006,
  title = {The Writhe of Open and Closed Curves},
  author = {Berger, Mitchell A and Prior, Chris},
  year = 2006,
  journal = {Journal of Physics A: Mathematical and General},
  volume = {39},
  number = {26},
  pages = {8321--8348},
  issn = {0305-4470, 1361-6447},
  doi = {10.1088/0305-4470/39/26/005},
  url = {https://iopscience.iop.org/article/10.1088/0305-4470/39/26/005},
}

@article{Panagiotou2013,
  title = {Writhe and Mutual Entanglement Combine to Give the Entanglement Length},
  author = {Panagiotou, E. and Kr{\"o}ger, M. and Millett, K. C.},
  year = 2013,
  month = dec,
  journal = {Physical Review E},
  volume = {88},
  number = {6},
  pages = {062604},
  issn = {1539-3755, 1550-2376},
  doi = {10.1103/PhysRevE.88.062604},
  url = {https://link.aps.org/doi/10.1103/PhysRevE.88.062604},
}

@article{Panagiotou2020,
  title = {Knot Polynomials of Open and Closed Curves},
  author = {Panagiotou, Eleni and Kauffman, Louis H.},
  year = 2020,
  month = aug,
  journal = {Proceedings of the Royal Society A: Mathematical, Physical and Engineering Sciences},
  volume = {476},
  number = {2240},
  pages = {20200124},
  issn = {1364-5021, 1471-2946},
  doi = {10.1098/rspa.2020.0124},
  url = {https://royalsocietypublishing.org/doi/10.1098/rspa.2020.0124},
}

@article{Orlandini1994,
  title = {The Writhe of a Self-Avoiding Walk},
  author = {Orlandini, E and Tesi, M C and Whittington, S G and Sumners, D W and Rensburg, E J Janse Van},
  year = 1994,
  journal = {Journal of Physics A: Mathematical and General},
  volume = {27},
  number = {10},
  pages = {L333-L338},
  issn = {0305-4470, 1361-6447},
  doi = {10.1088/0305-4470/27/10/006},
  url = {https://iopscience.iop.org/article/10.1088/0305-4470/27/10/006},
}

@article{Foteinopoulou2008,
  title = {Universal {{Scaling}}, {{Entanglements}}, and {{Knots}} of {{Model Chain Molecules}}},
  author = {Foteinopoulou, Katerina and Karayiannis, Nikos Ch. and Laso, Manuel and Kr{\"o}ger, Martin and Mansfield, Marc L.},
  year = 2008,
  month = dec,
  journal = {Physical Review Letters},
  volume = {101},
  number = {26},
  pages = {265702},
  issn = {0031-9007, 1079-7114},
  doi = {10.1103/PhysRevLett.101.265702},
  url = {https://link.aps.org/doi/10.1103/PhysRevLett.101.265702},
}

@book{Flapan2000,
  author    = {Flapan, Erica},
  title     = {When Topology Meets Chemistry: A Topological Look at Molecular Chirality},
  year      = {2000},
  publisher = {Cambridge University Press},
  address   = {Singapore},
}

@article{Laso2009,
  title = {Random Packing of Model Polymers: Local Structure, Topological Hindrance and Universal Scaling},
  shorttitle = {Random Packing of Model Polymers},
  author = {Laso, Manuel and Karayiannis, Nikos Ch. and Foteinopoulou, Katerina and Mansfield, Marc L. and Kr{\"o}ger, Martin},
  year = 2009,
  journal = {Soft Matter},
  volume = {5},
  number = {9},
  pages = {1762},
  issn = {1744-683X, 1744-6848},
  doi = {10.1039/b820264h},
  url = {https://xlink.rsc.org/?DOI=b820264h},
}

@article{Sulkowska2012,
  title = {Conservation of Complex Knotting and Slipknotting Patterns in Proteins},
  author = {Su{\l}kowska, Joanna I. and Rawdon, Eric J. and Millett, Kenneth C. and Onuchic, Jose N. and Stasiak, Andrzej},
  year = 2012,
  month = jun,
  journal = {Proceedings of the National Academy of Sciences},
  volume = {109},
  number = {26},
  issn = {0027-8424, 1091-6490},
  doi = {10.1073/pnas.1205918109},
  url = {https://pnas.org/doi/full/10.1073/pnas.1205918109},
}

@article{Shen2024,
  title = {Knot Data Analysis Using Multiscale {{Gauss}} Link Integral},
  author = {Shen, Li and Feng, Hongsong and Li, Fengling and Lei, Fengchun and Wu, Jie and Wei, Guo-Wei},
  year = 2024,
  month = oct,
  journal = {Proceedings of the National Academy of Sciences},
  volume = {121},
  number = {42},
  pages = {e2408431121},
  issn = {0027-8424, 1091-6490},
  doi = {10.1073/pnas.2408431121},
  url = {https://pnas.org/doi/10.1073/pnas.2408431121},
}

\end{document}